\UseRawInputEncoding
\documentclass[manuscript,screen]{acmart}
\usepackage{caption}
\usepackage{subfigure}
\usepackage{color}
\usepackage{longtable}
\usepackage{multirow}
\usepackage{amsmath}

\usepackage[english]{babel}
\usepackage{tikz}
\usetikzlibrary{positioning}

\newcommand{\eg}[1]{\textit{e.g.,}}

\AtBeginDocument{%
  \providecommand\BibTeX{{%
    \normalfont B\kern-0.5em{\scshape i\kern-0.25em b}\kern-0.8em\TeX}}}

\setcopyright{acmcopyright}
\copyrightyear{2023}
\acmYear{2023}
\acmDOI{XXXXXXX.XXXXXXX}

\begin{document}

\title{Stock Market Prediction via Deep Learning Techniques: A Survey}

\author{Jinan Zou}
\authornote{Both authors contributed equally to this research.}
\email{jinan.zou@adelaide.edu.au}
\author{Qingying Zhao}
\authornotemark[1]
\affiliation{%
  \institution{University of Adelaide}
  \country{Australia}
}

\author{Yang Jiao}
\affiliation{%
  \institution{University of Adelaide}
  \country{Australia}}

\author{Haiyao Cao}
\affiliation{%
  \institution{University of Adelaide}
  \country{Australia}
}

\author{Yanxi Liu}
\affiliation{%
 \institution{University of Adelaide}
 \country{Australia}}

\author{Qingsen Yan}
\affiliation{%
  \institution{Northwestern Polytechnical University}
  \country{China}}

\author{Ehsan Abbasnejad}
\affiliation{%
  \institution{University of Adelaide}
  \country{Australia}}
 
 \author{Lingqiao Liu}
\affiliation{%
  \institution{University of Adelaide}
  \country{Australia}}
\author{Javen Qinfeng Shi}
\affiliation{%
  \institution{University of Adelaide}
  \country{Australia}}

\renewcommand{\shortauthors}{Zou and Zhao, et al.}

\begin{abstract}

Existing surveys on stock market prediction often focus on traditional machine learning methods instead of deep learning methods.  This motivates us to provide a structured and comprehensive overview of the research on stock market prediction. We present four elaborated subtasks of stock market prediction and propose a novel taxonomy to summarize the state-of-the-art models based on deep neural networks. In addition, we also provide detailed statistics on the datasets and evaluation metrics commonly used in the stock market. Finally, we
point out several future directions by sharing some new perspectives 
on stock market prediction.

  \end{abstract}



\maketitle

\section{Introduction}

The financial market plays a crucial role in shaping the development of global economies. As the stock market gains more prominence in the economic sphere, it has garnered increasing attention from the general public. One theory that explains the pricing of financial assets and the reasoning behind stock market volatility is the efficient market hypothesis \citep{malkiel2003efficient}. The efficient market hypothesis posits that in a legally sound, well-functioning, transparent, and competitive stock market, rational investors can quickly and rationally react to all market information. As a result, stock prices will accurately, sufficiently, and promptly reflect all important facts, including a company's present and future value. However, the fluctuation of stock prices is influenced by a complex array of factors, including company earnings reports, national policies, influential shareholders, and expert speculations on current events. Therefore, there is a need to utilize machine learning techniques for stock market prediction tasks, such as stock movement prediction, stock price prediction, portfolio management, and trading strategies.

The stock market is characterized by both uncertainty and variability, making it challenging to accurately predict market trends. Machine learning techniques have been employed in stock price prediction to improve the accuracy of predictions and alleviate these difficulties. Historically, conventional models such as decision tree-based models \cite{vu2012experiment,nugroho2014decision,kamble2017short} and Support Vector Machines (SVM) \cite{xie2013semantic} have been utilized in stock market forecasting.

As deep learning models have evolved, the methods used for predicting the stock market have shifted from traditional techniques to advanced deep learning techniques such as Recurrent Neural Networks (RNNs), Long Short-Term Memory (LSTM), Gated Recurrent Units (GRUs), Graph Neural Networks (GNNs), and Convolutional Neural Networks (CNNs). Recently, researchers have also begun to explore the use of Transformer-based models and Reinforcement learning (RL) models in stock market forecasting. Despite the abundance of surveys on stock market forecasting, existing surveys \cite{atsalakis2009surveying, li2010applications, ballings2015evaluating, tkavc2016artificial} had limitations. For instance, some surveys have focused on traditional technologies without critically examining the latest advancements, such as Transformer models. Additionally, some surveys have been vague in their classification of models and have not used an authoritative criterion. Furthermore, many of the challenges and open issues identified in previous studies have already been addressed. This survey aims to fill these gaps by providing a comprehensive and insightful overview of the latest techniques and trends in stock market forecasting. By reviewing high-quality papers from top conferences, this survey summarizes the most recent advancements in techniques such as Transformer and RL and provides in-depth analysis and discussion. It is intended to provide researchers, practitioners, and educators with a systematic overview and a comprehensive understanding of relevant deep learning techniques and the most promising directions for future research.

This survey aims to gain insight into the advancements in stock market forecasting by categorizing the models and analyzing their publication year. Additionally, the survey aims to provide a detailed understanding of the structure and application of each model in stock market forecasting.

The three key contributions of this survey are as follows:
\begin{itemize}

\item 
In this survey, we thoroughly examine stock market prediction, which encompasses four distinct tasks: stock movement prediction, stock price prediction, portfolio management, and trading strategies. To conduct this study, we have compiled a collection of 94 papers that focus on these highly relevant topics.

\item
This survey introduces a new deep-learning classification system for predicting stock market performance. The reviewed literature, which is organized according to this taxonomy, explores various deep learning models such as RNN, CNN, GNN, Transformer, and RL. Furthermore, the survey compiles a summary of the datasets, evaluation techniques, and model inputs used in these studies.

\item 
In this study, we delve into the unresolved challenges faced by deep learning-based stock market prediction and offer thorough insights into potential future research in this area. 

\end{itemize}

\begin{figure}
\includegraphics[width=0.9\textwidth]{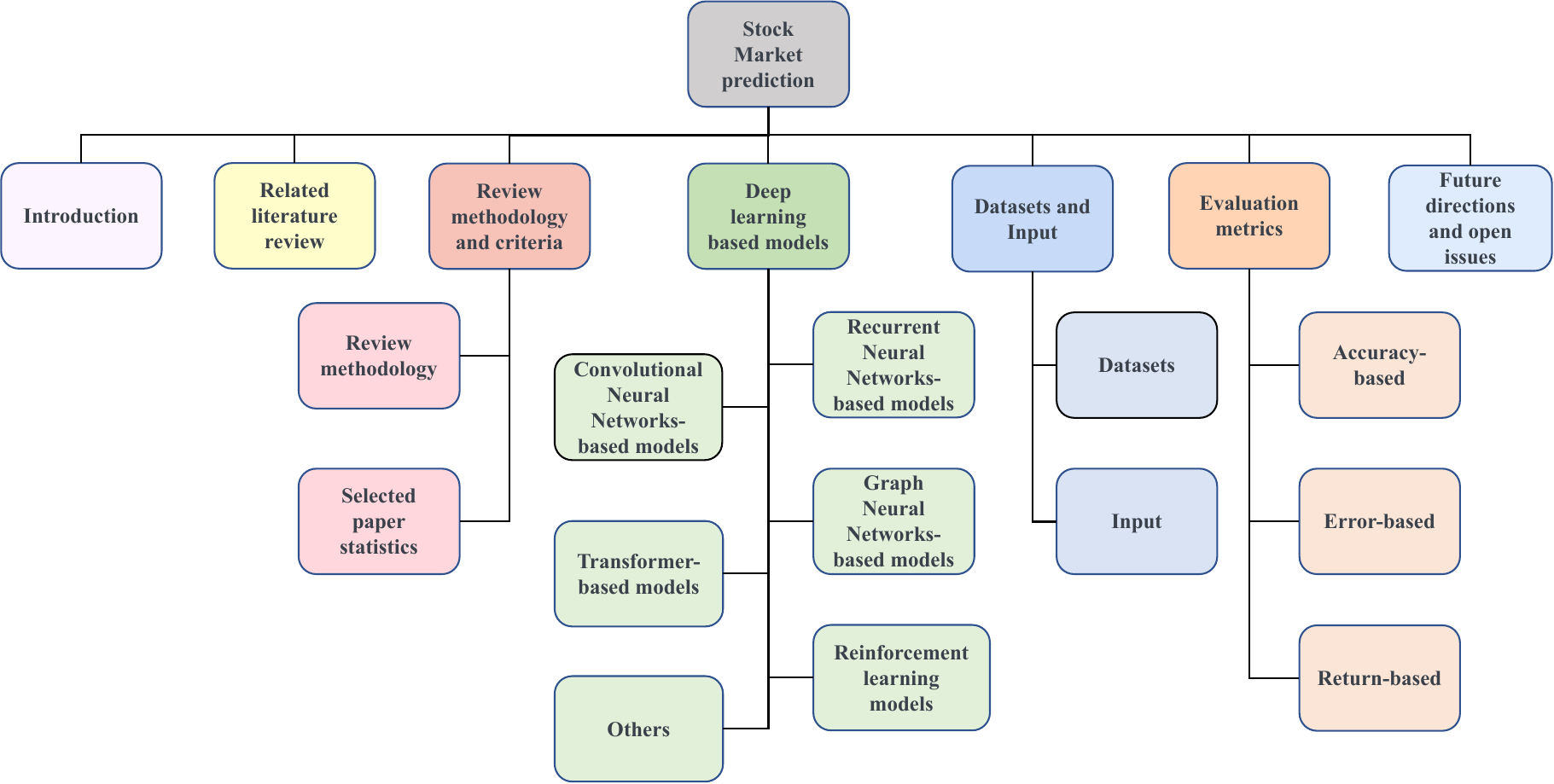}
\caption{The hierarchical structure of this survey with eight sections.} \label{fig1}
\end{figure}

\noindent\textbf {Organization of the Survey.} 
This survey is organized into eight sections. The hierarchical structure of the survey is illustrated in Figure \ref{fig1}. The first section provides an overview of the background, research motivation, and objectives of stock market forecasting. In Section \ref{Related work}, previous surveys on the topic are reviewed, and their limitations and areas for improvement are identified. Section \ref{review} details the methodology used for reviewing papers, including the criteria for categorization, such as conference, model, and year of publication. In Section \ref{model}, various deep learning models used in the stock market prediction are discussed, and the papers are analyzed based on model types. The datasets and model input features used in the reviewed methods are discussed in Section \ref{dataset}. Evaluation metrics commonly used in the stock market prediction are introduced in Section \ref{evaluation}. Section \ref{future} focuses on open issues and potential future developments in the field. The survey concludes with a summary in Section \ref{conclusion}.

\section{Related work}\label{Related work}

In past studies on stock prediction, traditional methods such as SVM, regression, and KNN have been widely used. However, with the advancement of neural networks, there has been a shift towards utilizing these networks in stock prediction research. Atsalakis and Valavanis \cite{atsalakis2009surveying} conducted a survey of 100 papers on stock market prediction, focusing on neural and neuro-fuzzy techniques. They classified and analyzed papers published from 1990 to 2006 by comparing different stock markets, input variables, prediction approaches, and evaluation metrics. Li and Ma \cite{li2010applications} published a survey on using artificial neural networks in financial market applications, including stock price prediction and option pricing. Ballings et al. \cite{ballings2015evaluating} focused on classifier models for ensemble methods in stock price prediction and single classifier models. The ensemble methods included Random Forest, Adaboost, and Kernel Factory, while the single classifier models included Neural Networks, Linear Regression, SVM, and K-Nearest Neighbors. This was the first study to make an extensive benchmark. Tkac and Verner \cite{tkavc2016artificial} provided a systematic overview of 412 papers published between 1994 and 2015 on the commercial application of artificial neural networks in the financial industry. The main applications of this survey included financial distress, bankruptcy problems, and decision support for stock price forecasting, with a focus on classification tasks. Cavalcante et al. \cite{cavalcante2016computational} reviewed influential papers using various machine learning algorithms from 2009 to 2015 in financial applications, and discussed challenges and unsolved problems in this field. The main machine learning methods covered included artificial neural networks, SVMs, hybrid methods, optimization methods, and ensemble methods. The applications included preprocessing and clustering financial data, forecasting future market trends, and mining financial text information. Li et al. \cite{li2017web} surveyed 229 papers from 2007 to 2016 investigating the relationship between web media and stock markets, and proposed directions for further research. They proposed a taxonomy that divided machine learning techniques into statistical models, regression models, and machine learning-based models. The neural network models included Bayesian classifiers and SVMs. Xing, Cambria, and Welsch \cite {xing2018natural} investigated papers on applying natural language processing methods to financial forecasting tasks, mainly focusing on three aspects: different types of text sources, algorithms, and results based on different evaluation metrics. The text sources included news, social media, financial reports, and message boards. The model types included regressions, SVM, ensemble learning, and CNN. Nti, Adekoya, and Weyori \cite {nti2020systematic} reviewed papers from 2007 to 2018, focusing on three categories of analysis: technical, fundamental, and combined analysis. The models were limited, and only three techniques were well discussed: decision tree, SVM and ANN. Ersan, Nishioka, and Scherp \cite {ersan2020comparison} also focused on three prediction models: KNN, ANN, and SVM. This survey aims to make comparisons among various machine learning approaches on DAX 30 and the S\&P 500 datasets at daily and hourly levels.

Recent surveys on stock market prediction have primarily focused on deep learning techniques such as CNN, RNN, and GNN. \citet{jiang2021applications} stood out among others as it took into consideration the implementation and reproducibility of the research. Specifically, the paper highlighted the main tools used for implementation, including Keras, TensorFlow, PyTorch, Theano, and scikit-learn. The data and code availability of several papers were also investigated to demonstrate reproducibility. This study placed emphasis on the recent advancements in deep learning methods applied to stock market prediction.
Thakkar and Chaudhari \cite{thakkar2021comprehensive} examined various deep learning-based neural network approaches for stock market prediction. The study summarizes and analyzes the need, challenges, and future directions based on papers from 2017 to 2020. Kumbure et al. \cite{kumbure2022machine} conducted a literature review of 138 journal articles on stock market forecasting published between 2000 and 2019. The survey places particular attention on the features and unique variables in the datasets. The survey categorizes the deep learning methods into two groups, supervised and unsupervised machine learning methods. However, some of the surveys still lack the inclusion of the latest techniques such as Transformer and pre-trained BERT models, which are also reviewed in this survey.

Previous surveys have extensively covered traditional methods and older neural network techniques, leaving a gap in understanding current research trends. Recent surveys have not been fully updated to include the latest techniques. Additionally, some existing surveys \cite{thakkar2021comprehensive, kumbure2022machine, jiang2021applications, li2017web, hu2021survey, kumar2021systematic} have not adequately classified papers by targets or have only focused on stock market forecasting tasks. Given this need for an updated survey that incorporates the latest techniques for stock market prediction, this survey proposes a new deep learning model classification method and aims to provide a more comprehensive and up-to-date analysis. The focus will be on newer papers and the most current technology.

\section{Review methodology and criteria}\label{review}
\begin{figure}
\includegraphics[width=1\textwidth]{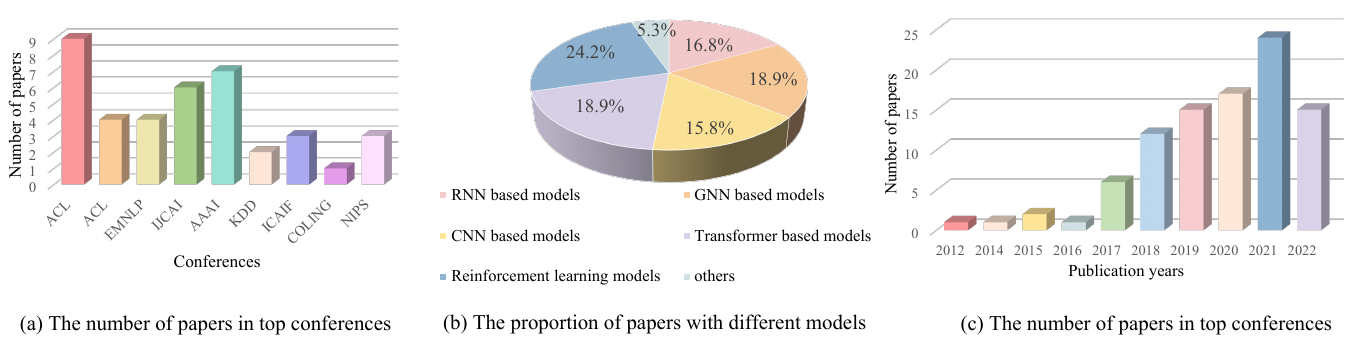}
\caption{Three Figures show the statistics of the papers according to different criteria.} \label{f1}
\end{figure}
\subsection{Review Methodology}
In this study, we focus on literature pertaining to stock market prediction. To gather a comprehensive collection of papers, we implemented a strategy as follows. Initially, we limited the timeframe to the last decade. Next, we identified the leading Natural Language Processing (NLP) and Artificial Intelligence (AI) conferences, including ACL, EMNLP, AAAI, IJCAI, ICAIF, NeurIPS, and KDD. Within these conferences, we utilized the search engine Google Scholar with specific keywords such as \textit{stock prediction}, \textit{market}, \textit{finance} and \textit{portfolio}. We also included the names of deep learning models as keywords, such as \textit{RNN}, \textit{LSTM}, and \textit{GNN}, \textit{Transformer}, and \textit{RL}. By using these keywords, we were able to find papers related to stock market prediction, finance, trading, and portfolio. We then used machine learning to filter papers that predicted the stock market. After filtering, we selected 42 high-quality papers related to stock prediction. After excluding papers with less than two pages of content, 39 papers were screened. By reading the cited papers in these 39 papers, we also included the top-ranked papers from journals and conferences in our research paper list. In total, this survey includes 94 high-quality papers that cover a diverse range of deep learning methods.

\subsection{Selected Paper Statistics}

This section presents statistics from the stock market forecast papers included in the survey through various charts. First, the papers are grouped by top conference. Figure \ref{f1}(a) displays the number of papers from top conferences, such as ACL\footnote{Annual Meeting of the Association for Computational Linguistics.}, NAACL\footnote{Conference of the North American Chapter of the Association for Computational Linguistics.}, EMNLP\footnote{Empirical Methods in Natural Language Processing.}, AAAI\footnote{Association for the Advancement of Artificial Intelligence.}, IJCAI\footnote{International Joint Conference on Artificial Intelligence.}, ICAIF\footnote{ACM International Conference on AI in Finance.}, NeurIPS\footnote{Conference and Workshop on Neural Information Processing Systems.}, ICML\footnote{International Conference on Machine Learning.}, ICLR\footnote{International Conference on Learning Representations.}, and KDD\footnote{Knowledge Discovery and Data Mining.}. As seen in Figure \ref{f1}(a), a total of 39 high-quality papers were included in this survey. The conferences of ACL and AAAI had the most papers on stock market prediction. Figure \ref{f1}(b) illustrates the proportion of papers utilizing five different models. This survey covers all the types of models mentioned, with each model having more than 15 papers. The survey includes advanced models, such as Transformer-based and RL models, which make up 18.9\% and 24.2\% respectively. Figure \ref{f1}(c) shows the number of papers by publication year. The largest proportion of papers in this survey were published in 2021, with a total of 24 papers. During the five-year period from 2012 to 2016, there were 11 papers.

\section{Stock market prediction in machine learning}\label{model}
\subsection{Stock Prediction Related Tasks}
Before delving into the specifics of the deep learning model, we will first define four key stock market prediction tasks and provide an overview of the concepts associated with each task. These tasks include stock price prediction, stock movement prediction, portfolio management, and trading strategies. These categories encapsulate the majority of existing stock market prediction tasks.

\begin{itemize}

\item 
\textbf{Stock Price Prediction.}
The objective of Stock Price Prediction utilizing time-series data is to anticipate future values for stocks and financial assets traded on an exchange by researchers. The ultimate goal of this prediction is to achieve substantial profits. Additionally, various factors also influence the prediction process, including psychological factors as well as rational and irrational behavior. All of these factors work together to make share prices dynamic and volatile.

\item 
\textbf{Stock Movement Prediction.}
The task of Stock Movement Prediction typically categorizes stock trends into three categories: uptrend, downtrend and sideways. This task is formalized by analyzing the difference between the adjusted closing prices of a stock over a certain period of trading days.

\item 
\textbf{Portfolio Management.}
Portfolio management involves the strategic selection and oversight of a collection of investments with the aim of achieving financial objectives.  The goal of portfolio management is to allocate resources in a way that maximizes returns while minimizing risks.
\item 
\textbf{Trading Strategies.}
A trading strategy is a pre-established set of guidelines and criteria that are used to make trading decisions, and it is a methodical approach to buying and selling stocks. Trading strategies can range from simple to complex, and factors such as investment style (e.g., value vs. growth), market cap, technical indicators, fundamental analysis, level of portfolio diversification, risk tolerance, and leverage are taken into consideration. In stock market prediction tasks that utilize deep learning, common trading strategies include event-driven, data-driven, and policy optimization.
\end{itemize}

\begin{figure*}[htbp]
\begin{center}
\includegraphics[width=0.95\textwidth]{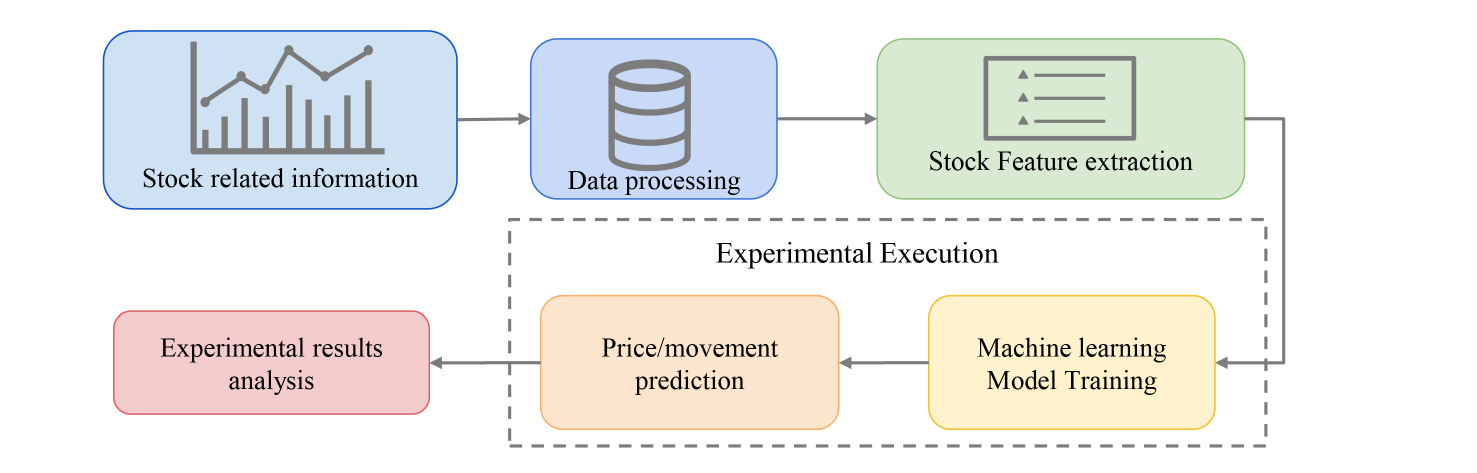}
\end{center}
\caption{The Processing Framework.} 
\label{Framework1}
\end{figure*}

The above tasks center around the process of stock market prediction. To provide a comprehensive understanding of deep learning-based approaches, Figure \ref{Framework1} illustrates the prediction process. The first step involves processing input data, including stock data, graphs, and texts. After that, relevant stock features are selected and collected. The next step is to input the extracted features into a deep learning model for training. Lastly, the model's experimental results will be obtained and analyzed.

\subsection{Deep Learning Based Models}

\begin{figure*}
 \center
  \includegraphics[width=0.8\textwidth]{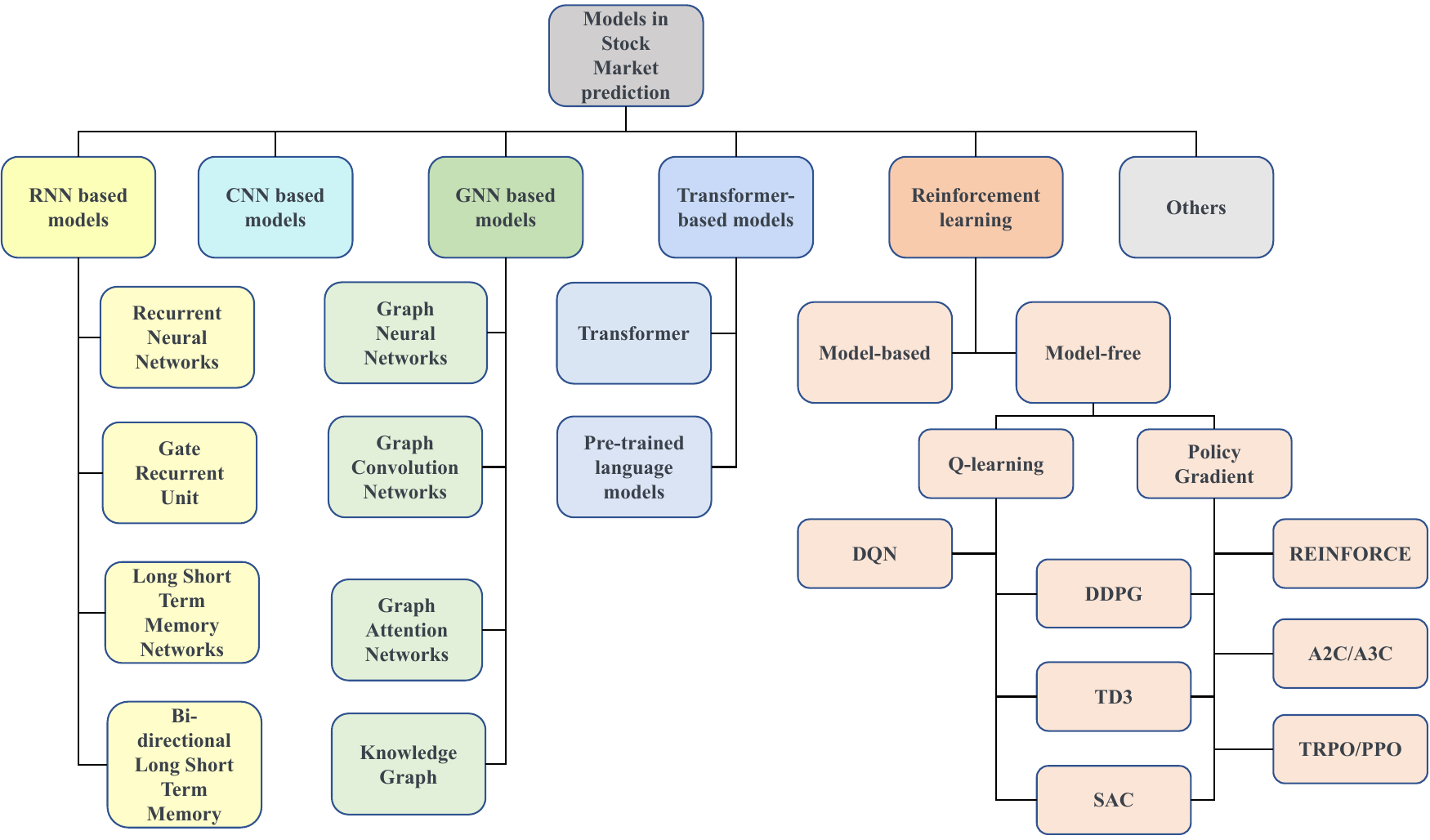}
  \caption{The classification method of stock market prediction papers according to the models.}
  \label{fig4}
\end{figure*}

\begin{figure*}[htbp]
\begin{center}
\includegraphics[width=0.98\textwidth]{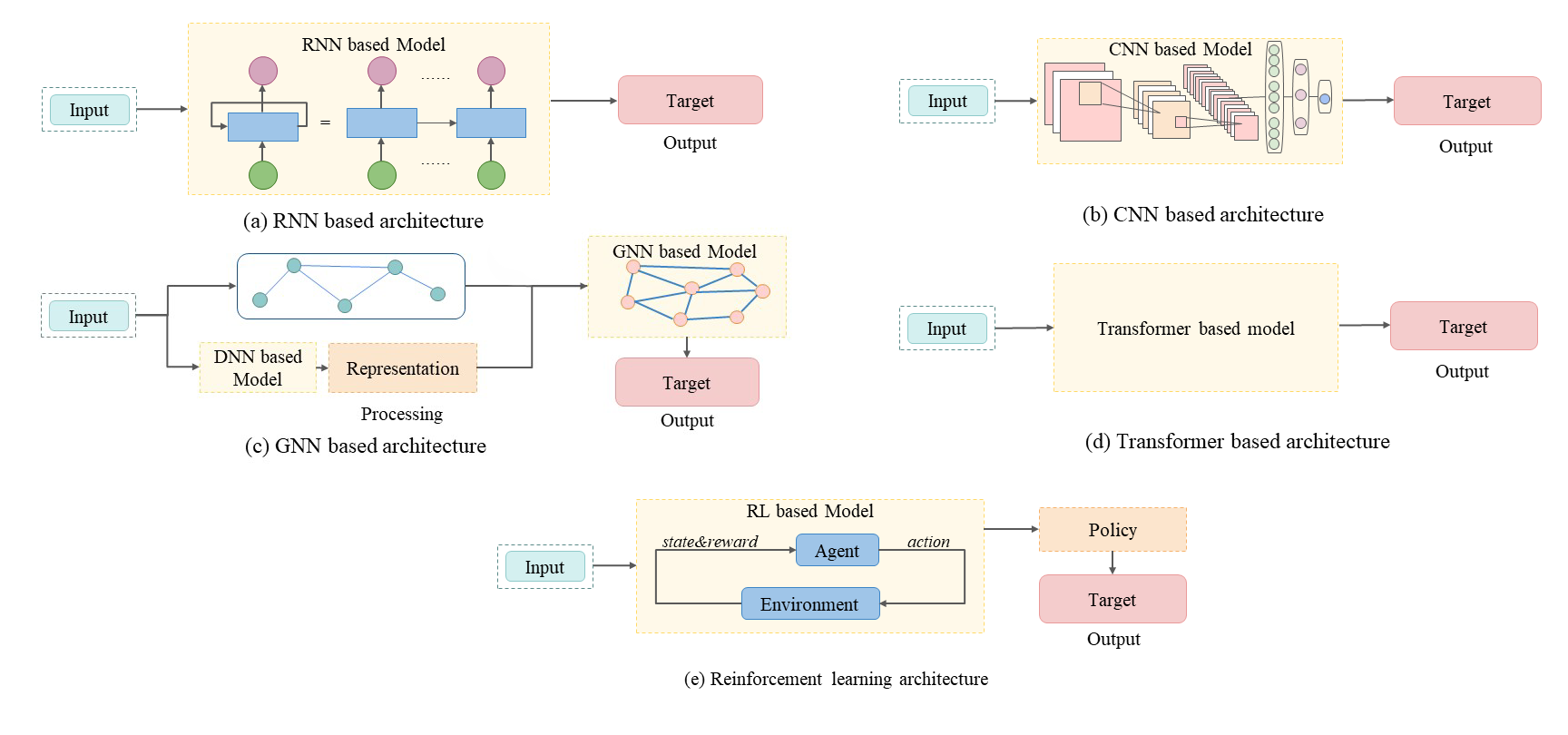}
\end{center}
\caption{The general view of mainstream deep learning models for stock market prediction.}
\label{Framework}
\end{figure*}

We categorized the papers based on the models and presented the different models in Figure \ref{fig4}. Figure \ref{Framework} provides a general overview of the mainstream deep learning models used in stock market prediction. These models include: RNN-based models (Figure \ref{Framework}(a)), CNN-based models (Figure \ref{Framework}(b)), GNN-based models (Figure \ref{Framework}(c)), RL models (Figure \ref{Framework}(d)), Transformer-based models (Figure \ref{Framework}(e)), as well as other unique methods. The input for these models can be stock price-related data, text data, and relationships between companies, and the models generate outputs based on the goals outlined in the papers.

\subsection{Recurrent Neural Network Based Models}

The RNN, as described in \cite{rumelhart1986learning}, is a deep learning model that can efficiently process sequential data. Stock market data is commonly represented as time series, making RNN an ideal choice for making predictions by modeling historical data. However, RNNs have the drawback of gradient vanishing when processing data over a long-term period. To address this issue, several variants of RNN have been developed, including LSTM \cite{hochreiter1997long}, GRU \cite{cho2014properties}, and Bidirectional LSTM (Bi-LSTM) \cite{graves2005framewise}. RNNs can be thought of as a recurrent combination of multiple identical network cells, where the output of each cell is provided as input to the next cell \citep{lei2021rnn}. Each cell contains a set of input, hidden, and output units. The LSTM unit has three gates, the update gate, the forget gate, and the output gate, which help to control short-term and long-term memory. The GRU model is an improvement over the LSTM model, which combines the input gate and the forget gate into an update gate, and adds a reset gate to control the information to be forgotten \citep{cho2014properties}. These models have improved RNNs and made significant progress in stock market prediction.

\noindent\textbf{Recurrent Neural Network (RNN).} RNN is a well-established deep learning model that has been applied in stock prediction. However, in recent years, researchers have sought to enhance the performance of RNN by exploring its hybrid application with other machine learning techniques. Agarwal and Sastry  \cite{rather2015recurrent} proposed a novel and robust hybrid prediction model (HPM) that combines three prediction models: RNN, Exponential smooth (ES) and Autoregressive moving average model (ARMA). They also used genetic algorithms to optimize the model by providing optimal weights which significantly improves the prediction accuracy. Another approach, proposed by Zhang, Aggarwal and Qi \cite{zhang2017stock}, is the State Frequency Memory (SFM) based on RNN, which is able to capture multi-frequency trading patterns from the stock market underlying the fluctuation of stock prices.

\noindent\textbf{Long Short-Term Memory (LSTM).} The LSTM model boasts the ability to effectively handle text and time-series data, making it ideal for stock market forecasting. LSTM addresses the problem of preserving information over longer time intervals through the use of the gradient method, which is an improvement over RNN models. For example, Akita et al. \cite{akita2016deep} proposed a Paragraph Vector method to represent textual information and utilized LSTM for the prediction model. In this experiment, ten companies were represented as ten articles, where the vector of articles is represented as $P_{t}$. The prices of these companies were represented as $\{c1, c2…, c10\}$ at a single timestep $t$ and concatenated as a stock prices vector $N_{t}$. The input of LSTM is the combination of $P_{t}$ and $N_{t}$. LSTM significantly outperforms the baselines, Multi-Layer Perceptron (MLP), Support Vector Regression (SVR), and Simple-RNN, for the opening prices prediction. Another example, Ma et al. \cite{ma2019news2vec} proposed the News2vec model, in which dense vectors represent news features. They used LSTM with the self-attention mechanism as the predictive model. This text embedding model News2vec helps to uncover the potential relationship between news and its event elements. Nelson, Pereira and Oliveira \cite{nelson2017stock} used LSTM for stock movement prediction, but the model inputs were numerical information, including stock prices and volume.

In stock prediction research, it is common to process stock prices as a time series analysis, but few researchers take into account the potential time dependency of the data. Zhao et al. \cite{zhao2017time} proposed a time-weighted LSTM model with trend retracement, which assigns weights to data based on its temporal proximity to the data to be predicted. Hybrid models, which are a combination of LSTM with other models, have been proposed to improve prediction performance. Polamuri et al. \cite{polamuri2021multi} proposed the Generative Adversarial Network-based Hybrid Prediction Algorithm (GAN-HPA) to implement the Stock-GAN developed from a GAN-based framework. The framework takes various inputs, such as stock datasets and hyperparameters. Linear and non-linear models are used to extract features from the dataset, and the hyperparameters and pre-processing results are used as inputs to the LSTM. The output of the generator and raw data are provided to the discriminator, and Bayesian approximation is used to adjust the parameters and update the prediction results. Wang et al. \cite{wang2021clvsa} made a significant contribution by proposing a method for dynamically extracting potential representations of financial market trends from transaction data. They proposed a hybrid Convolutional LSTM-based Variational Sequence-to-Sequence model with Attention (CLVSA). The model consists of convolutional LSTM units and a sequence-to-sequence framework with self-attention and inter-attention mechanisms.

Nguyen and Yoon \cite{nguyen2019novel} developed a framework called deep transfer with related stock information (DTRSI) that takes into account stock relationships to predict stock price movements. To achieve this, the authors employed LSTM cells during pre-training on large-scale data to optimize parameters and then fine-tuned the base model using a small amount of target data to obtain the final model. This approach addresses the problem of overfitting due to the small sample size and accounts for the relationship between stocks.

Adversarial training can simulate stock price fluctuations by introducing perturbations, which can enhance the accuracy of stock movement prediction. In 2019, Feng et al. \cite{feng2018enhancing} proposed a method that combines attentive LSTM and adversarial training to predict stock market movements. Another approach to predicting multiple stock prices simultaneously is the associated network model proposed by Ding and Qin \cite{ding2020study}, which consists of three branches that predict the opening price, the lowest price, and the highest price respectively.

Chen et al. \cite{chen2019incorporating}  used Bi-LSTM to encode stock data and financial news representations in their models, namely the Structured Stock Prediction Model (SSPM) and multi-task Structured Stock Prediction Model (MSSPM). They also incorporated representations of trading events consisting of event roles embeddings, which are the subject and object extracted using the $Standford$ $CoreNLP^2$ tool \cite{rather2015recurrent}. The MSSPM model was uniquely trained on both stock prediction and event extraction tasks, as the authors believed these tasks have an intrinsic relationship, and accurate extraction results would benefit stock prediction results. To make time-aware predictions, Sawhney et al. \cite{sawhney2021fast} proposed a hierarchical learning method named FAST for ranking stocks based on expected profit. This model used a time-aware LSTM to model temporal irregularities in news and tweets. The FAST model demonstrates the positive effects of factoring in-text fine-grained temporal irregularities in simulations on  the S\&P 500 and China A-shares indexes.

Li and Pan \citep{li2022novel} proposed that stock prices are impacted by a variety of factors and put forth an ensemble deep learning model that combines LSTM and GRU. They also determined the window size for textual information by consulting psychological and economic theories to account for the lasting or fleeting impact of news.

\noindent\textbf{Gated Recurrent Unit (GRU).} GRUs are a variation of LSTMs that have been found to perform well in stock market prediction. They address the issue of vanishing gradients and improve training speed by reducing the number of cells \cite{cho2014properties}.  Inspired by the work of Qin et al. \cite{qin2017dual}, who proposed a dual-stage attention-based RNN for time series prediction, Yang et al. \cite{yang2018explainable} proposed a dual-level attention mechanism for stock price prediction using a GRU network. This mechanism allocates different weights to different financial news titles based on their impact on the stock price, with a focus on explaining the reasons for the prediction and avoiding errors from natural language processing tools. Similarly, Li, Shen, and Zhu \cite{li2018stock} proposed a novel multi-input LSTM (MI-LSTM) model that uses attention to differentiate between main and auxiliary factors and assign different weights to these inputs to prevent the influence of irrelevant factors on the final result.

To address the challenge of chaotic news, Hu et al. \cite{hu2018listening} developed a Hybrid Attention Network (HAN) that incorporates a self-paced learning mechanism. This approach takes into account both the credibility and comprehensiveness of financial news. The HAN consists of two attention layers, one at the news level and the other at the temporal level. Bi-directional GRU layers are also utilized to encode the temporal sequence of corpus vectors. After back-testing, the annualized rate of return in simulated transactions was significantly improved.
Similarly, Wang et al. \cite{wang2020incorporating} proposed a framework for the stock prediction that incorporates expert opinions. The multi-view fusion network stance detection model (MFN) framework uses text features from multiple views and relevant financial domain knowledge to determine texts' upside and downside investment opinions. A stance aggregation module is then used to identify and aggregate high-quality opinions based on a dynamic expert mining procedure. Finally, a stock prediction module is used to predict the future trend of individual stocks using the input expert opinion indicators from the first two sections. The prediction component utilizes a GRU to encode a time-ordered sequence of features and a time-aware attention mechanism to dynamically incorporate hidden states in the stocks. The model was validated using real-world datasets, demonstrating its effectiveness in making investment recommendations and individual stock predictions.

In an effort to tackle the difficulties of stock trend forecasting caused by non-smooth dynamics and complex market dependencies, Wang et al. \cite{wang2021hierarchical} introduced a new Hierarchical Adaptive Temporal-Relational Network (HATR) for describing and predicting the evolution of stocks in 2021. The HATR model captures both short-term and long-term transition characteristics from multi-scale local combinations of stock trading series through the use of expanding causal convolutions and gating paths. This model builds upon their previous work, in which they utilized a dual attention mechanism with a Hawkes process and target-specific queries to detect key time points and scales based on individual stock characteristics. Additionally, to uncover hidden interdependencies among stocks, the authors incorporated a multi-graph interaction module that combines prior domain knowledge with data-driven adaptive learning.

\subsection{Convolutional Neural Network Based Models}
CNNs have been widely studied for their effectiveness in both computer vision (CV) and natural language processing (NLP) tasks \cite{ismail2019deep}. A CNN model is composed of several convolutional and pooling layers that are used for feature extraction. The traditional convolutional layers use two-dimensional filters (kernels) and activation functions to process image features. However, in the stock prediction domain, CNNs are used to process time series data, which are one-dimensional features. To accommodate for this difference in data shape, CNNs for time series utilize a one-dimensional filter that slides over the time series with a stride determined by the data granularity.

It is commonly believed that CNNs excel at capturing important features through the use of convolution layers, making them effective at predicting stock fluctuations. Selvin et al. \cite{selvin2017stock} supported this view. Additionally, Ding et al. \cite{ding2015deep} improved upon the event embedding method \cite{ding2014using} by incorporating a CNN for training on input history event embeddings, with the pooling layer effectively extracting representative event history features. The Universal CNN-based predictor (U-CNN pred) \cite{hoseinzade2019u} was trained using a layer-wise approach, where the subCNNs layers were pre-trained sequentially until the proposed model structure was completed. This model showed reasonable results and was proven effective due to its shallow structure, which reduced the risk of overfitting as there were less weights to be learned.

To fully utilize the information contained within industrial relations, some models have integrated knowledge graphs with CNNs in order to enhance their performance. One such model is the Knowledge-Driven Temporal Convolutional Network (KDTCN) proposed by Deng et al. \cite{deng2019knowledge}. This model utilizes the Open IE \cite{etzioni2008open} to extract events related to the knowledge graph and make interpretable stock predictions. A common issue with traditional one-dimensional convolutional layers is data leakage, where information from time $t-1$ and $t+1$ could influence the data at time $t$. To address this issue, the KDTCN model employs causal convolution, which only used information from the current and previous time steps in the previous layer.
The KDTCN model has been shown to be effective in explaining abrupt price changes by extracting significant features in the price time series \cite{deng2019knowledge}.

The integration of CNN and LSTM can enhance time series prediction even further. Lu et al. \cite{lu2020cnn} introduced a CNN-LSTM model for forecasting daily stock closing prices, where the CNN component extracts features from a 10-day historical data time series, and the LSTM component makes the price prediction. In a subsequent study, Lu et al. \cite{lu2021cnn} proposed a CNN-BiLSTM-AM model which incorporates the attention mechanism to capture historical influential stock fluctuations on price time series and improve the performance of the CNN-based model. Wang et al. \cite{wang2021stock} also presented a CNN-BiLSTM model for stock closing price prediction and improved the model's performance by adding a tanh function to the output gate of the Bi-LSTM. Mehtab and Sen \cite{mehtab2020stock} proposed a univariable convolutional LSTM model for predicting the opening price of Indian stocks and improved the model's performance by partitioning a 10-day time series into two 5-day data sequences to allow the convolutional layer to extract more historical data features. Additionally, the use of GRU has also been recently demonstrated as efficient in processing tasks. Zhou, Zhou, and Wang \cite{zhou2022stock} proposed an ensemble stock market prediction model consisting of CNN and Bidirectional GRU, which is feature-selection-based. The CNN is responsible for feature extraction, while the GRU is responsible for processing the time series data. They used the closing price of the stock market as the model output and all other data as inputs and obtained results with less error than other basic models.

Some recent studies have explored the integration of knowledge graphs, LSTM, and CNN for stock prediction. Wu et al. \cite{wu2020cnn} developed a CNN-based framework that utilized historical stock prices and future data, such as leading financial indicators. In their subsequent work \cite{wu2021graph}, they proposed a graph-based CNN-LSTM model that uses a combination image composed of an option image, a future image, and a historical image. This image contains 30-day information, including stock price and financial indices of one specific stock. The rows of the combination image represent the time series changes, while the columns represent the features that are fed into the CNN-LSTM model. To address the issues of overfitting and slow convergence in fuzzy systems, Chandar \cite{chandar2022convolutional} developed a robust stock trading model. This model extracted ten technical indicators from historical stock data and used them as feature vectors. By setting the data range, the convergence rate was improved. These feature vectors were then transformed into images and used as input for a CNN model to obtain labelled sell points, buy points, and hold points. The dropout layer added in the CNN models helped to prevent overfitting. The effectiveness of this model was measured by accuracy and F1-score.

\subsection{Graph Neural Network Based Models}

GNN is a type of artificial neural network that process data in the form of graphs, as outlined in \citep{scarselli2008graph}. They play a crucial role in stock market prediction, as they are able to operate on irregularly structured data, unlike CNNs which are designed for Euclidean structured data. The structure of a GNN is made up of nodes and edges, which allows it to model the relationships between entities. In the context of stock market prediction, nodes typically represent companies or stocks and edges represent the relationships between them. For example, the share prices of linked companies often fluctuate simultaneously, such as when a piece of good news is released and corresponding stocks see an immediate surge. This highlights the importance of taking relationships into account when making predictions. This chapter will explore four main graph-based models: GNN, Graph Convolutional Networks (GCN) \citep{kipf2016semi}, and Graph Attention Networks (GAT).

\noindent\textbf{Graph Neural Network (GNN).} Matsunaga, Suzumura and Takahashi \cite{matsunaga2019exploring} employed a knowledge graph to integrate information on companies and GNN models as a means of predicting individual stock performance. The use of knowledge graphs allows for the representation of relationships between entities that represent companies in the context of stock market predictions. One of the key contributions of the paper is the utilization of a backtesting approach through rolling window analysis. Similarly, Ding et al. \cite{ding2016knowledge} built upon their previous event-driven works \cite{ding2014using,ding2015deep} by introducing a Knowledge Graph Neural Tensor Network (NTN) model. This model encodes the entity vectors that represent the relationship between entities of extracted events and feeds them into the event-embedding learning process, addressing the limitation of event embedding in revealing syntactic or semantic relationships between events.

Xu et al. \cite{xu2022hgnn} tackled the challenge of stock price limit prediction by introducing the Hierarchical Graph Neural Network (HGNN). This model takes into account the various properties of the market state by constructing a stock market relationship graph and extracting information from multiple perspectives, such as the node view, relation view, and graph view in a hierarchical manner. The HGNN achieved excellent results in classifying the type of price-limit-hitting stocks and resulted in an improvement in the investment return ratio.
Similarly, Li et al. \cite{li2022graph} proposed a GNN-based model for predicting the stock market by fusing multi-source heterogeneous subgraphs. The datasets used in this research included three types of subgraphs representing the relationships between the stock market index, stock market news, and graphical indicators. The fusion of these subgraphs was then converted into a fully connected classification layer to make predictions. Ang and Lim \cite{xu2022hgnn} utilized a graph encoding module to propagate multimodal information across companies' relationships. Additionally, they introduced an attention module for capturing global and local information among inter-company relations and different modalities. The model performed robustly on three forecasting tasks and two applications on real-world datasets.

\noindent\textbf{Graph Convolutional Network (GCN).}
GCN is a type of deep learning model that are specifically designed to handle graph data, which uses graph convolutional layers to extract features from the graph and make predictions based on the relationships between nodes in the graph. GCN is often combined with other deep learning models. For example, Chen and Wei \cite{chen2018incorporating} proposed a pipeline prediction model that integrates the relationships between corporations by using a GCN model. In this model, each corporation is represented as a node in the graph, with edges representing the relationships between corporations and the weights of those edges representing the shareholding ratios. Additionally, the LSTM-based encoder layer is used to encode historical features of corporations, resulting in improved performance. Similarly, Li et al. \cite{li2021modeling} proposed an LSTM Relational Graph Convolutional Network (LSTM-RGCN) model that handles both positive and negative correlations among stocks. The correlation matrix between companies is calculated based on historical market data, and the LSTM mechanism added to the RGCN layers helps to alleviate the over-smoothing problem when predicting overnight stock price movements. A novel Gated temporal convolution is also introduced to learn the temporal evolution of stock features.

Several existing models aimed to capture the temporal dependence between stock prices and news information. However, they did not fully utilize information from other highly correlated stocks. To address this gap, Yin et al. \cite{9533510} introduced a graph convolutional network model that integrates GCN and GRU. GCN extracts features from stocks that have a high degree of similarity, which are then fed into the GRU model to capture the time dependence. Another approach to consider the relationship between stocks is the Relational Stock Ranking (RSR) framework proposed by Feng et al. \cite{feng2019temporal}. The RSR framework consists of three layers: a sequential embedding layer using LSTM, a relational embedding layer, and a prediction layer. Additionally, they proposed a temporal graph convolution model to solve the ranking problem. Sawhney et al. proposed the Spatio-Temporal Hypergraph Convolution Network (STHGCN) \cite{sawhney2020spatiotemporal}, a notable approach that used a hypergraph structure to model relationships among stocks and applies spatial hypergraph convolutions. This was the first hypergraph learning approach.

Predicting the price movements of individual stocks can be a difficult task due to various factors such as company operations and public opinion. However, stock market indices provide a more reliable means of understanding the overall trend of a specific industry or company in the stock market as they are less affected by the variables of a single company. In a recent study, Wang et al. \cite{WANG2022108285} employed the use of a GCN to analyze the correlation of indicators in stock trend prediction. They introduced the MG-Conv model, which is based on a multi-Graph Convolutional neural network and utilizes static graphs between indices that are constructed using constituent stock data. Additionally, they created dynamic graphs based on trend correlations between indices with different portfolio strategies and defined multi-graph convolution operations based on both graphs.

\noindent\textbf{Graph Attention Networks (GAT).}
The Graph Attention Network (GAT) combines the strengths of GNN and attention layers to improve performance in large-scale graphs. Attention mechanisms help to focus on the most critical nodes, reducing the impact of complex background noise and improving the signal-to-noise ratio. Additionally, attention allows for exploiting interconnectedness and hierarchical connections among nodes to enhance the relevant information for the task at hand. Kim et al. \cite{kim2019hats} proposed using a hierarchical attention network (HATs) to predict individual stock prices and market index movements using relational data. HATs employ LSTM and GRU as feature extraction modules for these respective tasks and achieve better results than previous methods by aggregating different types of data and adding this information to each representation. Similarly, Sawhney et al. \cite{sawhney2020deep} proposed a multipronged attention network for stock forecasting (MAN-SF) that fuses information from financial data, social media, and inter-stock relationships using hierarchical attention to train a GAT.

One approach for forecasting a specific firm's trend is using GCNs based on pre-established relationships among firms. However, momentum spillovers can occur through various firm connections, the significance of which can change over time. Cheng and Li \cite{cheng2021modeling} introduced an attribute-driven graph attention network (AD-GAT) to capture these attribute-driven momentum spillovers. This network utilizes an unmasked attention mechanism to infer the dynamic firm relationships within the market signal, utilizing a feature extraction module based on tensors. The proposed model was found to be more accurate and have a higher AUC than GCN, eLSTM \cite{8966989}, and TGC \cite{feng2019temporal} in experiments using three years of data from the S\&P 500.

\subsection{Transformer Based Models}
CNNs excel in handling spatial data by creating an internal representation of two-dimensional information. In addition, RNNs are better suited for tasks involving temporal or sequential data, such as financial news, tweets, and stock price time series. However, RNNs may struggle with processing long sequences as the model can forget the contents of distant locations or mix up the contents of nearby positions. The Transformer addresses this issue by utilizing a self-attention mechanism and positional embedding to process sentences. As a result, the Transformer model has shown promising results in various stock market prediction tasks.

\noindent\textbf{Transformer Based Models.}
In an effort to enhance stock volatility models, Ramos-Pérez et al. \cite{ramos2021multi} implemented machine learning and deep learning techniques. They proposed the Multi-Transformer model, a variation of the existing Transformer model, which utilizes a strategy of randomly selecting various training data subsets and incorporating multi-attention methods to increase the stability and accuracy of attention processes. Similarly, Ding et al. \cite{ding2020hierarchical} introduced the hierarchical multi-Scale Gaussian Transformer for predicting stock movements. They improved upon the traditional Transformer by incorporating multi-scale Gaussian prior and optimizing locality, as well as implementing Orthogonal Regularization to prevent redundant learning heads under multiple attention. Additionally, they developed the Trading Gap Splitter for the Transformer to aid in learning the structural hierarchy of high-frequency financial data.

Transformer models have been shown to effectively capture long-term dependencies, making them well-suited for tackling problems related to temporal dependence. Li et al. \cite{li2022incorporating} proposed a novel Transformer encoder attention (TEA) framework that utilizes attention mechanisms to address issues of time dependency in financial data and reveal hidden information in stock prices related to social media texts. The TEA model utilizes a feature extractor and cascade processor architecture, consisting of a Transformer encoder, attention mechanism, and normalization technique. The feature extractor effectively gathers information from past text and stock prices for five calendar days in order to extract crucial information. Similarly, Zhang et al. \cite{ZHANG2022117239} introduced the Transformer-based attention network (TEANet) architecture to handle time-dependent problems utilizing five calendar-day data. The TEANet framework includes a deep textual feature extractor that utilizes the Transformer and a concatenation processor to effectively incorporate and balance the influence of various elements, such as tweets and market prices. Yoo et al. \cite{10.1145/3447548.3467297} improved predictive accuracy by leveraging connections between multiple equities. To achieve this, they introduced the Data-axis Transformer with Multi-Level Contexts (DTML). The DTML model builds asymmetric and dynamic correlations in an end-to-end approach to learn the correlations between stocks and provide final predictions for all individual stocks.

Many studies utilizing Transformer-based models employed textual information as input to understand the sentiment in stock-related news media. The goal of financial news sentiment analysis is to predict market reactions to the underlying information in texts \cite{yang2020finbert}. According to Li et al. \cite{li2017role}, social sentiment plays a crucial role in reflecting public views on stock trends. To gather this information, a collection of social sentiments and professional opinions was collected from both social platforms and financial news articles. This data was then fed into a tensor Transformer for model training, which helped to eliminate noise and capture more intrinsic relationships. The trained model could then be used to explore the impact and function of social emotions using data from various sources. Liu et al. \cite{liu-etal-2019-Transformer} argued that existing social media-based stock prediction algorithms only considered individual stock semantics and correlations, but failed to account for the contradicting information present on vast social media platforms. They proposed a Capsule network based on Transformer Encoder (CapTE) as a solution, which includes a Transformer Encoder to capture deep semantic features and structured relationships among tweets. Yang et al. \cite{10.1145/3366423.3380128} proposed a Hierarchical, Transformer-based, multi-task (HTML) model for predicting short-term and long-term asset volatility. Additionally, they used audio data in addition to common news and reports about finance to make predictions.

Chen et al. \cite{chen_chen_shen_shi_wang_zhang_2022} introduced the Gated Three-Tower Transformer (GT3) as a solution for extracting and integrating multivariate stock time series. To address the challenge of limited receptive fields, they implemented the Shifted Window Tower Encoder (CWTE) for capturing channel-wise features from data embedding. In order to extract and aggregate multi-scale temporal information, the team developed the Shifted Window Tower Encoder (SWTE) with multi-temporal aggregation. For sophisticated text feature extraction, the team employed a vanilla Transformer encoder as the Text Tower Encoder (TTE). Additionally, the Cross-Tower Attention method was implemented to aid the model in understanding market tendencies and the meanings conveyed by social media content. The features from CWTE, SWTE, and TTE are then fused through an adaptive gate layer for efficient and accurate results.

\noindent\textbf{Pre-trained Language Model.} BERT, a language model based on the Transformer architecture, has become a popular choice for pre-training in natural language processing tasks \cite{devlin-etal-2019-bert}. The model utilized two unique training methods, namely masked-language modeling (MLM) and next sentence prediction (NSP) \cite{devlin-etal-2019-bert}, to gain an understanding of relationships between words and long-term dependencies between sentences. Additionally, BERT's pre-trained model could be fine-tuned to suit specific use cases.

Financial news is considered a key source of information for stock market analysis and its impact on stock returns has been well-documented \cite{LI201414}. Dong et al. \cite{9378345} proposed a BERT-LSTM model that uses the BERT to extract the direction of stock prices based on social media news, while the autoregressive LSTM integrates information features as covariates. The model also utilizes historical price trends to predict future stock price movements. Sonkiya et al. \cite{sonkiya2021stock} employed the BERT model for sentiment analysis on news and headlines about Apple Incorporation. The sentiment scores obtained from the analysis were used as input vectors for a GAN, which consisted of a GRU and a CNN as the generator and discriminator. The GAN was able to generate data continuously and discriminate between true and generated samples of stock prices, thus achieving the desired prediction effect. The model's early convergence was optimized by using sentiment scores as input. Colasanto et al. \cite{COLASANTO2022341} improved stock forecasting by utilizing AlBERTo \cite{Polignano2019AlBERToIB}, a Transformer-based model, for sentiment analysis of Italian social media. This model calculates the sentiment values of various event news in the market that may affect stocks.

Instead of relying solely on the sentiment found in texts for stock market prediction, some researchers propose that news comments can influence investors' sentiment and ultimately affect their estimation of market trends and investment decisions. Li et al. \cite{li_li_wang_jia_rui_2020} used the BERT pre-training model to evaluate and classify investor comments found on news websites. They applied a cross-sectional regression analysis approach to validate the relationship between investor emotions and stock returns, utilizing a two-step cross-sectional regression validation \cite{RePEc:ucp:jpolec:v:81:y:1973:i:3:p:607-36} to eliminate any potential issues with heteroskedasticity and consistency in the data. Zhao et al. \cite{ZHAO2022117958} also acknowledged the importance of expert stock commentary for accurate stock prediction and thus chose to utilize BERT for more comprehensive and accurate translations of comments from field experts. They noted that the fixed-length text input of BERT can lead to poor performance in exploring long text information. To overcome this limitation, they employed a sliding window technique to segment the original text, increasing the sample size and reducing over-fitting, to capture all information from the lengthy text. Furthermore, they extracted the output features from each layer of the BERT model and applied an ablation strategy to extract useful information from these features.

The utilization of BERT in the stock market is not limited to just predicting prices or movements. Zhou et al. \cite{zhou2021trade} proposed a bi-level BERT-based model for detecting predefined trading events, which is further enhanced by incorporating a wide range of financial texts. The low-level model is a multi-label token classifier that identifies events for each token in each phrase. The high-level model combines the output of the low-level model with the entire article to determine the likelihood of each event occurring. The final trading strategy is based on the recognized time and ticker, utilizing string matching to detect events. Hsu et al. \cite{hsu-etal-2021-semantics} adopted a selective perturbed masking (SPM) approach for aspect-based sentiment analysis. SPM analyzes the value of each word in a sentence and replaces insignificant words using two replacement strategies without compromising aspect-level polarity to tackle readability and semantic consistency issues. The authors tested SPM for stock price and risk change prediction as a real-world scenario for sentiment analysis and further evaluated it in sub-tasks such as aspect term sentiment classification (ATSC) and aspect term extraction (ATE).

\subsection{Reinforcement Learning Models}
\label{RL}
\begin{figure}
\centering
\includegraphics[width=1\textwidth]{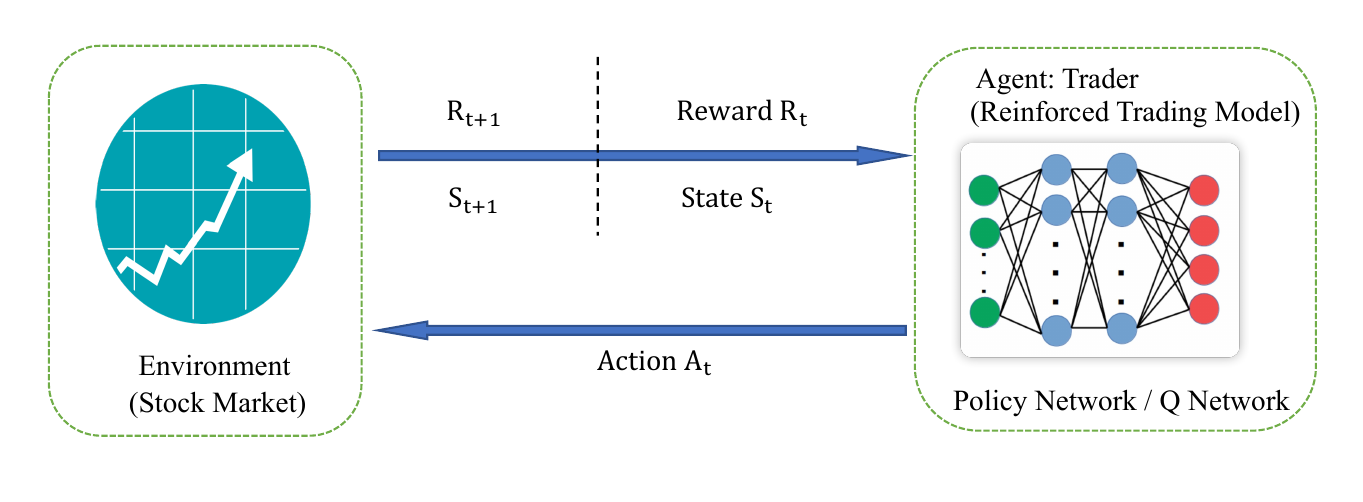}
\caption{In the field of financial trading, the interaction between the agent (trader) and the environment (financial market) is crucial. RL algorithms utilize this interaction by considering the financial market and trader as the environment and agent, respectively. Within the agent, different RL algorithms may incorporate the use of policy networks and Q networks. The financial data and returns in finance can be viewed as the state $S$ and reward $R$ in RL, while the trading transaction can be considered as the action $A$ in RL.}
\label{fig:mdp}
\end{figure}
In the stock market, RL is utilized to design trading strategies and manage portfolios. RL is a framework that allows for learning through interactions with the environment, as depicted in Figure \ref{fig:mdp}. Key concepts in RL include the Markov Decision Process (MDP) \citep{puterman2014markov}, agent, environment, and reward signal. The RL problem can be formulated as follows: the agent optimizes its policy through interactions with the environment. Specifically, the agent consists of a state and a policy represented by $S_t$ and $\pi$ at time t. When the agent interacts with the environment, a reward $r$ is received and the agent's state is updated to $S_{t+1}$. A decision process is considered Markov if the next state is solely dependent on the current state.

MDP also can be defined in the form of a tuple: 
$
    \mathcal{M} = \{\mathcal{S}, \mathcal{A},\mathcal{R}, \mathcal{T}, \gamma \},
$
where the states and actions are represented by $\mathcal{S} \in \mathbb{R}^n$ and $\mathcal{A} \in \mathbb{R}^m$, respectively. The function $\mathcal{R}: \mathcal{S} \times \mathcal{A} \rightarrow \mathbb{R}$ represents the reward, and the function $\mathcal{T}: \mathcal{S} \times \mathcal{A} \rightarrow \mathcal{S}$ represents the next state. The discount factor $\gamma$ is used to decrease the impact of future rewards. The objective of RL is to optimize the policy, in order to maximize the expected return.

The general trading process, which includes both high-frequency trading and portfolio management, can be visualized in Figure \ref{fig:mdp}. This process can be formalized as a decision-making process, represented by the MDP $\mathcal{M} = {\mathcal{S}, \mathcal{A},\mathcal{R}, \mathcal{T}, \gamma }$. In this MDP, the state $s$ contains fundamental information and market data for companies, as well as the position held by the agent. The action $a_t$ is a continuous or discrete vector indicating the number of stocks being traded at a given time step. The reward $r$ for each time step is the profit gained at that step. The state transition $\mathcal{T}$ is determined by both the state and action. The goal of RL algorithms is to find a policy that maximizes the expected return $R = \underset{\tau}{E} [ \sum_{i=1}^{\infty} \gamma^{t-1}r_t]$ over a trajectory $\tau$.

\subsubsection{Model-free Reinforcement Learning}
Model-free RL algorithms are a well-established branch developed in recent decades where the agent interacts directly with the environment. In model-free settings, policy gradient, Q-learning, and hybrid algorithms are widely used in the financial market.

\noindent\textbf{Policy Gradient.} The RL aims to maximize the expected return. One direct way is to maximize the objective function $J(\pi_\theta)= E_{\tau\sim\pi_\theta}[R(\tau)]$. The gradient of this function, known as $\nabla J(\pi_\theta)$, is called the policy gradient. By using the Log-Derivative Trick, the gradient can be transformed into:

\begin{equation}
\nabla_{\theta} J\left(\pi_{\theta}\right)=\underset{\tau \sim \pi_{\theta}}{\mathrm{E}}\left[\sum_{t=0}^{T} \nabla_{\theta} \log \pi_{\theta}\left(a_{t} \mid s_{t}\right) R(\tau)\right].
\end{equation}

The REINFORCE algorithm \citep{williams1992simple} is the basic policy gradient algorithm. The Advantage Actor-Critic (A2C) \citep{mnih2016asynchronous} algorithm improves REINFORCE by adding a baseline to reduce the variance and a critic to estimate the state values, which can evaluate the action token. To improve the efficiency of A2C, the Asynchronous Advantage Actor-Critic (A3C) \citep{mnih2016asynchronous} algorithm was proposed, which uses multiple agents to train the actor-critic network asynchronously. The Trust Region Policy Optimization (TRPO) \citep{schulman2015trust} and Proximal Policy Optimization Algorithms (PPO) \citep{schulman2017proximal} set constraints on the KL-Divergence and clip, respectively, on how much difference there is between the old policy and the new policy and take the biggest improvement step under these constraints.

The REINFORCE algorithm has gained attention in the field of financial trading due to its ability to be optimized through gradient ascent. For instance, Liang et al. \cite{liang2018adversarial} evaluated the effectiveness of three different RL algorithms, including DDPG, PPO, and REINFORCE, in an adversarial training setting and found that REINFORCE performed the best. These experiments were conducted on the China stock market and the authors suggested that the policy gradient-based method is particularly well-suited for financial scenarios. Additionally, incorporating historical information into the state by using a policy gradient with a recurrent network has also been a promising approach. Jiang, Xu, and Liang \cite{jiang2017deep} developed a model-free financial RL framework for portfolio management that incorporates CNN, RNN, and LSTM, and was built on the deterministic policy gradient (DPG). The framework was tested in the cryptocurrency market and was shown to outperform other methods.

The actor-critic-based method has been shown to be effective in reducing policy variance by incorporating state value. Li, Rao, and Shi \citep{li2018learning} and Ponomarev, Oseledetsa, and Cichocki \citep{ponomarev2019using} both employed actor-critic methods (A2C and A3C respectively) in combination with CNNs to analyze the China stock market and RTS index future respectively. In their study, Li, Rao, and Shi \citep{li2018learning} found that their proposed Deep Actor-Critic Trading (DACT) strategy outperformed other methods such as buying and holding, DQN, and REINFORCE. Meanwhile, Ponomarev, Oseledetsa, and Cichocki \citep{ponomarev2019using} achieved promising results, achieving 66\% profitability per year even when accounting for commission fees.

Several studies have sought to improve performance in quantitative trading by utilizing information on the relationships between different stocks, rather than solely focusing on improving RL algorithms. For instance, Wang et al. \cite{wang2019alphastock} developed the AlphaStock method, which utilizes a sharp ratio-oriented policy gradient approach to address challenges in portfolio management such as balancing profit and risk and avoiding extreme losses. The method fuses information on the relationships between assets and is the first to implement an interpretable trading strategy using Deep Reinforcement Learning (DRL). The algorithm was tested in both the US and China stock markets, and the results showed its effectiveness, robustness, and generalization ability. It tends to select stocks that have increasing tendencies and low volatility. Building on AlphaStock, Wang et al. \cite{wang2021deeptrader} proposed a new policy gradient trading algorithm called DeepTrader for portfolio management. It includes an Asset Scoring Unit (ASU) and a Market Scoring Unit (MSU). The ASU extracts temporal and spatial features crossing assets using dilated CNN and attention mechanisms, respectively, and also introduces GCNs to represent interrelationships and causal relationships between assets. The MSU controls overall positions for long and short, and its main network is an LSTM that extracts historical market representation. DeepTrader was tested in different stock markets and achieved the most profitable performance with less risk compared to baselines.

\noindent\textbf{Q-learning method.} Q-learning is a method for obtaining the optimal policy by updating the action-state value, represented by $Q$. The Bellman update is used to update $Q$ until it converges to the optimal value. The optimal policy can then be found by searching the $Q$ values greedily. The emergence of the Deep Q-learning Network (DQN) brought Q-learning to a new level by using a deep neural network to approximate the Q value. Prior to DQN, many studies in finance trading utilized tabular Q-learning algorithms \citep{dempster2001computational,jangmin2006adaptive,chen2007genetic,lee2007multiagent,bertoluzzo2012testing,eilers2014intelligent}. DQN-based models have shown promising results in this field. Carapucco Neves and Horta \citep{carapucco2018Reinforcement} used a DQN-based RL agent for forex trading, resulting in stable learning and environmental construction. Lucarelli and Borrotti \cite{lucarelli2019deep} proposed a Double Dueling DQN for cryptocurrency trading and evaluated the performance of different reward functions, such as the profit reward function and sharp ratio reward function. Theatre and Ernst \citep{theate2021application} developed a DQN-based trading algorithm using a limited stock market dataset and introduced new assessment indicators, including the Sharp ratio, Sortino Ratio, and profit or loss ratio.

Researchers have been increasingly drawn to variants of DQN with recurrent networks due to their ability to handle time-series data. Huang \cite{huang2018financial} proposed an MDP model for stock trading that combines time, market, and position features, and leveraged a recurrent DQN with small replay memory and action augmentation to solve the model. Chen and Gao \cite{chen2019application} also used a variant of DQN, the Deep Recurrent Q-network (DRQN), for automated trading on S\&P 500 ETF data and achieved superior performance compared to other methods. Additionally, Tsantekidis et al. \citep{tsantekidis2020price} applied recurrent DQN in the forex market with the use of a novel reward function that incorporates price trailing, profit, Sharpe ratio, and maximum drawdown to enhance the algorithm's simplicity and effectiveness.

Recurrent algorithms, such as RNN, LSTM and GRU, that predict market behaviour often rely solely on price information. However, as Carta et al. \cite{carta2021multi} have noted, using only price information and a supervised method to predict future market behaviour is highly challenging. To address this issue, the authors proposed a multi-layer and ensemble-based trading agent utilizing the DQN. This algorithm utilizes different meta-learners to maximize rewards and generate trading signals in various iterations. Additionally, all the learners are fused to make the final decision for trading. The proposed algorithm was tested in both future and stock markets and was found to outperform considered baselines. 

\noindent\textbf{Hybrid methods.}
Hybrid RL algorithms simultaneously learn both the policy and Q function. One such example is the Deterministic Policy Gradient method, which includes variations such as Deep Deterministic Policy Gradient (DDPG) \citep{lillicrap2015continuous}, Twin Delay DDPG (TD3) \citep{fujimoto2018addressing}, and Soft Actor-Critic (SAC) \citep{haarnoja2018soft}. In particular, DDPG utilizes an off-policy approach to learn both the Q value and policy from the Q function. TD3 incorporates techniques such as Clipped Double-Q Learning, Delay Policy Updates, and Target Policy Smoothing to prevent the overestimation of Q values. SAC, on the other hand, combines entropy regularization with Clipped Double-Q Learning and updates the Q-network using Polyak averaging.

The use of DDPG in the financial market has been widely studied by researchers due to its ability to combine the strengths of policy gradient and Q-learning in continuous action spaces. Xiong et al. \cite{xiong2018practical} applied DDPG algorithms to a custom environment consisting of 30 Dow Jones Industrial Average component stocks in order to develop a stock strategy. The agent's performance was evaluated by comparing it to the Dow Jones Industrial Average index and a traditional min-variance portfolio allocation strategy. The results revealed that the proposed agent demonstrated exceptional profitability.

In a similar vein, Bao and Liu \citep{bao2019multi} employed DDPG with multiple agents to tackle the Liquidation problem. This approach allows agents to both compete and cooperate with one another, resulting in a significant improvement over single-agent algorithms. Additionally, Sawhney et al. \cite{sawhney2021quantitative} introduced the PROFIT model, a deep RL-based approach that utilizes time-aware text analysis to model market information and optimize trading actions. This model was found to surpass competing methods in terms of profitability and risk management on the S\&P 500 and China A-shares indexes.

RL algorithms have the potential to be used in stock trading, however, creating and training a practical RL algorithm can be challenging and prone to errors. To address this issue, Liu et al. developed the FinRL library \cite{liu2020finrl}, which focuses on model-free RL algorithms and simplifies the implementation process for users. The library includes tutorials and various financial environments such as NASDAQ-100, DJIA, S\&P 500, HSI, SSE 50, and CSI 300, as well as popular DRL algorithms like DQN, DDPG, PPO, SAC, A2C, and TD3. Additionally, FinRL includes backtesting metrics for fair and objective evaluation. The authors has continued to improve the platform with updates such as \citep{liu2021finrl,li2021finrl,liu2021finrlmeta}.

Researchers have continued to investigate alternative approaches to RL despite its advancements in the field. Liu et al. \cite{liu2020adaptive} identified two major challenges in the application of RL to quantitative trading: handling noisy high-frequency financial data and balancing exploration and exploitation. To address these challenges, they proposed the adaptive trading method iRDPG, which combines imitation learning with the Recurrent Deterministic Policy Gradient. This approach treated the trading process as a partially observable MDP and was trained using minute-level data, highlighting its robustness and adaptability in different markets. Similarly, Wang et al. \cite{wang2021commission} recognized that existing methods were impractical due to their disregard for price slippage. To address this issue, they proposed a hierarchical model that incorporated a Reinforcement Learning algorithm for maximum return at the lower level and a policy gradient algorithm for portfolio weight generation at the higher level. The low-level RL also utilizes one-step temporal-difference learning to minimize trading costs and make quick trades.

\subsubsection{Model-based Reinforcement Learning}
In recent years, model-based algorithms have gained increased attention in the field of RL. These algorithms utilize a parameterized approximator, denoted as $\hat{p}_\eta$, to simulate the dynamics of a MDP defined by the set of states and actions $\mathcal{S,A}$. The model allows for the approximation of both the reward and the next state, given a specific state and action. Compared to model-free RL algorithms, model-based methods offer several advantages. Firstly, they can improve data efficiency by addressing challenges related to real-world data sampling, such as time-consuming or hardware-sensitive processes. Secondly, the parameterized model allows for interesting exploration strategies. Additionally, model-based methods can improve performance by combining local planning and global learning. Furthermore, dynamic models can be deployed in new tasks, making them useful in transfer learning scenarios. Additionally, the model's ability to capture causality can aid in solving both the intervention problem, which addresses the impact of a specific action and the counterfactual problem, which addresses the potential outcome of different actions in a specific situation.

Model-based RL has been used in the development of several trading algorithms due to its potential to improve the performance of dynamics transition models. Researchers, such as Yang, Yu, and Almahdi \citep{yang2018investor}, have noted that investor sentiment plays a significant role in the market and have therefore sought to design trading systems that incorporate this sentiment using the Gaussian inverse RL method. This method aimed to uncover the intrinsic mapping of investors' sentiment to market conditions and predict future market trends. While not a model-based method, the research by \citet{yang2018investor} provided insight into the potential for using such methods in trading. Wei et al. \citep{wei2019model} took a different approach by training a world model using Limit Order Book data with a frequency of once per 0.17s. By using a dynamics transition model, the RL agent can interact with a simulated world for optimization instead of the real environment. The authors also claimed that the trading policy trained using this model can be transferred directly to the real environment with stable profitability. Model-based RL has also been applied to portfolio management tasks. Yu et al. \citep{yu2019model} proposed an architecture that includes a prediction model, a generative adversarial data augmentation model, and a policy cloning model. When combined with the DDPG algorithm and trained on hourly price data, the model demonstrated both profitability and robustness. Briola et al. \cite{briola2021deep} also employed a model-based approach by building an end-to-end DRL agent using the PPO algorithm. The agent learns the transition dynamics and implements planning to achieve long-term returns. The algorithm uses limited order book data and selects training samples with significant price changes. The results showed that the proposed agent could produce stable profits in non-stationary markets.

\subsection{Other Deep Learning Methods}

In the field of event-driven stock forecasting, there are two key issues related to the use of indicator data sources: (1) the low reliability of individual sources and (2) the lack of understanding of the interactions and correlations among multiple sources. To address these challenges, Zhang et al. \cite{zhang2018improving} developed a coupled matrix and tensor factorization approach. This approach involves the creation of a quantitative feature matrix, the construction of a matrix, the extraction of events and sentiments, and the application of coupled matrix and tensor factorization. The resulting model is able to effectively fill in missing values in sparse tensors, allowing for accurate forecasting of market movements through the use of factorized low-rank matrices.

Stock movement prediction is challenging in highly stochastic stock markets. To solve this problem, Xu and Cohen \cite{xu2018stock} have proposed a solution through their novel deep generative model, Stocknet. This model utilizes both text and price signals from Twitter data and the previous five days' price data to predict the stock movement on the sixth day. The Stocknet model is composed of three parts: the Market Information Encoder (MIE), which encodes tweets and prices; the Variational Movement Decoder (VMD), which decodes stock movements; and the Attentive Temporal Auxiliary (ATA), which integrates temporal losses through an attention mechanism.

To address the issue of inadequate generalization caused by uncertainty in data and models, Wang et al. \cite{wang2021coupling} introduced a copula-based contrastive predictive coding (Co-CPC) method. Co-CPC considers the dependencies between a stock class, sector, and related macroeconomic variables, and learns stock representations from a micro perspective in a self-supervised manner. This allows for the mapping of stock characteristics to a generalized embedding space. The system combines microstock context with diverse macroeconomic factors and captures the coupling through a self-supervised objective of minimizing uncertainty in data and the model. Similarly, Duan et al. \cite{duan2018learning} proposed a novel target-specific abstract-guided news document representation model for extracting the most informative content. The model uses a target-sensitive representation of the news abstract to weigh sentences in the news content, allowing for the selection and combination of the most informative sentences for market modelling.

The factor model is a commonly employed asset pricing model in quantitative investment strategies. One of the major obstacles in constructing efficient factor models is the low signal-to-noise ratio present in financial data. To address this issue, Duan et al. \citep{duan2022factorvae} introduced the FactorVAE, which combines a dynamic factor model with the use of a variational autoencoder for noise modeling. By approximating the factor posterior factor model with future information, the FactorVAE can effectively guide the learning process.

\section{Dataset and Model Input}\label{dataset}

In the field of stock prediction, the datasets utilized by machine learning models vary depending on the perspective being taken in different stock markets. Two tasks that draw the most attention are predicting stock prices and price movements. Additionally, much of the research utilizing reinforcement learning (RL) centers around developing trading policies. When it comes to the input features used in the models, datasets can be broadly classified into two groups: intrinsic and extrinsic data. Intrinsic data primarily includes information extracted from the stock data itself, such as historical stock prices, financial indexes, and other technical analysis data. As stock data is inherently time-series in nature, intrinsic data is typically composed of time-series data. Moreover, extrinsic data can be quite varied and may include information such as text, fundamental data, industrial knowledge graphs, and more \cite{jiang2021applications}. Furthermore, the datasets used in this area of research can cover a wide range of time periods, from a few months to a decade or more. Another notable characteristic of the datasets is that they often come from different regions, with the majority being based on the US market, but also including markets from China, Japan, and India. A list of commonly used stock market abbreviations with country information can be found in Table \ref{tab: abbreviation}. Table \ref{tab: dataset} provides an overview of the datasets and model details used in the reviewed papers.

\begin{table*}[!htp]\centering
\caption{Introduced Stock Market Abbreviations.}\label{tab: abbreviation}
\small
\begin{tabular}{l|l|l}\hline
Abbreviation &Country &Full Name \\
\hline
CSI-300 &China &Index of top 300 stocks in Shanghai and Shenzhen stock exchanges \\
SSE &China &Shanghai Stock Exchange \\
SZI &China &Shenzhen Component Index \\
SZSW &China &Shenzhen Stock Exchange \\
HKEX &China &Honkong Stock Exchange \\
NSE &India &National stock exchange of India \\
Nifty 50 &India &The benchmark index of the NSE (National Stock Exchange) of India \\
BSE &India &Bombay stock exchange \\
Nikkei 225 &Japan &225 companies in Tokyo Stock Exchange \\
TOPIX &Japan &Tokyo Stock Price Index \\
KOSPI &Korea &The Korea Composite Stock Price Index \\
S\&P 500 &US &Standard and Poor 500 Index \\
SPX &US &S\&P 500 index \\
DJIA &US &Dow Jones Industrial Average \\
NASDAQ &US &National Association of Securities Dealers Automated Quotations Stock Market \\
NYSE &US &New York Stock Exchange \\
IWD &US &iShares Russell 1000 Value \\
IWC &US &iShares Micro-Cap \\
SPY &US &SPDR S\&P 500 ETF \\
DEM &US &WisdomTree Emerging Markets High Dividend \\
VTI &US &Vanguard Total Stock Market ETF \\
\hline
\end{tabular}
\end{table*}

\begin{enumerate}
    \item Stock Price. The stock price is considered the most direct reflection of the stock market's performance and is commonly used as both an input feature and a predicted target in various models. In the papers reviewed in this survey, the use of stock price was prevalent, but it was utilized in various forms such as open, high, low, and closed data, depending on the model design.
    \item Technical analysis tools. Technical analysis tools, commonly employed in traditional stock analysis, have a strong correlation with stock market performance. These tools take into account factors such as exchange rate, book-market ratio, trading volume, and other relevant financial indicators.
    \item Macroeconomic data. Macroeconomic data reflects the economic status of a specific region. Two commonly used indicators that are linked to the stock market are the Consumer Price Index (CPI) and Gross Domestic Product (GDP). These indexes provide insight into the current market conditions and indicate whether the stock market is experiencing growth or decline \cite{jiang2021applications}. 
    \item Fundamental data. Fundamental data refers to comprehensive information about an economic entity, including financial conditions, corporate structure, and any other information shared with shareholders. However, when it comes to utilizing this data in deep learning models, only a small portion is utilized due to limitations such as low reporting frequency and unstructured textual information.
    \item Knowledge graph. Different industries may have intrinsic connections, such as corporations in a single supply chain being affected by the same news. Recent experiments have shown that incorporating knowledge graphs from open sources with traditional stock data can improve the performance of models.
    \item Textual information. Textual information encompasses a wide range of sources, including but not limited to news articles, reports, social media posts, and user comments. Given that a majority of this information is unstructured, sentiment analysis is a widely employed technique for extracting insights through deep machine learning. The data can be categorized into various classes, such as positive, neutral, or negative, for further analysis and utilization.
\end{enumerate}

\subsection{Input Features}
The input features are extracted and organized based on the prediction target and dataset constitution, which can be roughly divided into four groups: time series, texts, knowledge graphs, and others.
\begin{enumerate}
    \item Time series. Time series data is a prevalent input in stock prediction due to the fact that many models rely on modeling stock price over time. The specific time frame of the prediction, such as intraday or interday, can determine the granularity of the data used, ranging from minute-level to day-level. Additionally, in the context of reinforcement learning, the time series data can be transformed into an environment where features can be utilized for creating states and rewards. This allows agents to interact with the environment and continually improve their decision-making policies.
    \item Text. Textual information encompasses a broad range of information sources such as news and articles. This type of information is thought to have a ripple effect on investor emotions. However, before being utilized in models, the textual information must undergo preprocessing and structuring, as it may originate from various languages and sources.
    \item Graph. The industrial knowledge graph is widely utilized, not only for displaying direct connections between corporations but also for uncovering internal relationships such as upstream and downstream supply chains.
    \item Others. Different data sources have been utilized in the stock prediction task, each offering unique perspectives. These include image data and audio data \cite{10.1145/3366423.3380128}. These data are employed as supplementary information, for instance, vocal features, such as voice tones, which can indicate the sentiment of speakers.
\end{enumerate}

\begin{center}
\fontsize{6}{8}\selectfont{
\begin{longtable}{p{1.5cm}|p{1.5cm}|p{1.8cm}|p{2.2cm}|p{4.5cm}|p{1.5cm}}

\caption{Datasets and models in reviewed papers. The column 'Input' refers to the model input features, while 'Target' refers to the model prediction target.} \label{tab: dataset}\\


\toprule
Author &Methods &Input &Target &Dataset &Time Span\\
\midrule
\endfirsthead


\multicolumn{6}{l}{Continued}\\
\toprule
Author &Methods &Input &Target &Dataset &Time Span\\
\midrule
\endhead

\bottomrule
\endfoot

\bottomrule
\endlastfoot

Rather \cite{rather2015recurrent} 
&RNN &Price &Stock Return &NSE &2007 - 2010 \\
Akita \cite{rather2015recurrent}
&LSTM &Price, Text &Stock Price &Nikkei 225 &2001 - 2008 \\
Nelson \cite{akita2016deep}
&LSTM &Price &Price Movement &Brazilian stock exchange &2014 \\
Zhao \cite{zhao2017time} &LSTM &Price &Price Movement &Yahoo! Finance. &2002 - 2007 \\
Zhang \cite{zhang2017stock} &RNN &Price &Price Movement &Yahoo! Finance &2007 - 2016 \\
Yang \cite{yang2018explainable} &GRU &Text, S\&P 500 &Price Movement &Reuters, Bloomberg &2006 - 2013 \\
Li \cite{li2018stock} &LSTM &Price &Stock Price &CSI-300 index &2013 - 2017 \\
Hu \cite{hu2018listening} &GRU &Price, News &Price Movement &Chinese stock data, Economic news &2014 - 2017 \\
Nguyen \cite{nguyen2019novel} &LSTM &Stock pice &Price Movement &KOSPI 200 and the S\&P 500 &2012 - 2018 \\
Wang \cite{wang2021clvsa} &LSTM &Price, Indicators &Price Movement &Oil, Gold, Gas, Soybeans, S\&P 500 and Nasdaq 100 &2010 - 2017 \\
Feng \cite{feng2018enhancing} &LSTM &Price &Price Movement &NASDAQ, NYSE &2007 - 2017 \\
Ma  \cite{ma2019news2vec} &LSTM &Price, News &Price Movement &Sohu, SSE &2009 - 2016 \\
Chen  \cite{chen2019incorporating} &Bi-LSTM &Event, Price, News &Price Movement &finance news related to TOPIX top 1000 stocks &2011 - 2017 \\
Ding  \cite{ding2020study} &LSTM &Price &Stock Price &SSE, PetroChina on SSE, ZTE on SZI. & \\
Shen  \cite{shen2012stock} &Hypergraph &Price, Indicator &Price Movement &SSE &2008 \\
Luo \cite{luo2014stock} &Hypergraph &Price &Price Movement &SSE &2008 \\
Chen \cite{chen2018incorporating} &GCN &Stock data, Graph &Stock Price &CSI 300 &2017 - 2017 \\
Kim \cite{kim2019hats} &GAT &Stock data, Graph &Price Movement &U.S. stock market, related Wikidata &2013 - 2019 \\
Feng \cite{feng2019temporal} &GCN &Stock data, Graph &Return Ratio &NASDAQ, NYSE &2013 - 2017 \\
Matsunaga \cite{matsunaga2019exploring} &GNN &Stock data, Graph &Stock Price &Nikkei 225 &2009 - 2019 \\
Sawhney \cite{sawhney2020spatiotemporal} &GCN &Price, Graph &Price Movement &S\&P 500, Yahoo finance &2013 - 2019 \\
Sawhney \cite{sawhney2020deep} &GAT &Price, Graph, Text &Price Movement &S\&P 500, NYSE，NASDAQ, Yahoo Finance. &2014 - 2016 \\
Li  \cite{li2021modeling} &GCN &Price, Graph, Text &Price Movement &TPX500, TPX100 index, Reuters Financial News &2018 - 2018 \\
Wang  \cite{wang2020incorporating} &MFN &Price, Text &Stock Price &Guba &2017 - 2018 \\
Wang  \cite{wang2021hierarchical} &
Conv., Gating
&Price, Text &Price Movement &CSI-300, S\&P 500, TOPIX-100 &2015-2020 \\
Xu \cite{xu2022hgnn} &HGNN &Stock data, Graph &Price-limit-hitting Stocks &SSE, SZSE &2018 - 2019 \\
Li  \cite{li2022graph} &GNN &Stock, Text, Graphical Indicators &Price Movement &SSE,CSI,SZI &2013 - 2019 \\
Xu  \cite{xu2021hist} &HIST &Stock Data &Price Movement &CSI 100, CSI 300 &2007 - 2020 \\
Ding  \cite{ding2015deep} &NTN &Price, Text &Price, Index &S\&P 500, related news &1999 \\
Deng  \cite{deng2019knowledge} &KDTCN &DJIA, Text &Price Movement &DJIA, Reddit news, Freebase, Wikidata &2008 - 2016 \\
Rasheed  \cite{rasheed2020improving} &CNN &Price &Stock Price &Amerisource Bergen Corporation, Cardinal Health & \\
Eapen  \cite{eapen2019novel} &CNN &Price &Stock Price &S\&P 500 grand challenge dataset on Yahoo &2008 - 2018 \\
Wu \cite{wu2020cnn} &CNN &Price, Indicators &Price Movement &TSM, NYSE, NASDAQ & \\
Lu \cite{lu2021cnn} &CNN-BiLSTM-AM &Price &Stock Price &SCI &1991 - 2020 \\
Lu \cite{lu2020cnn} &CNN-LSTM &Price &Stock Price &SCI &1991 - 2020 \\
Wu \cite{wu2021graph} &SACLSTM &Price &Stock Price &Stock price of USA and Taiwan & \\
Wang  \cite{wang2021stock} &CNN-BiSLSTM &Price &Stock Price &SZCI &1991 - 2020 \\
Mehtab \cite{mehtab2020stock} &CNN-LSTM &Stock index, Price &Stock Price &NIFTY 50 &2009 - 2020 \\
Hoseinzade \cite{hoseinzade2019u} &U-CNNpred &Stock index, Price &Price Movement &S\&P 500 &2010 - 2017 \\
Zhou \cite{zhou2021trade} &BERT &Price, Text &Stock Events &EDT dataset, Financial articles, S\&P 500 index, ETF &2020 - 2021 \\
Ding \cite{ding2020hierarchical} &Transformer &Price, Indicator &Price Movement &NASDAQ, CSI-500 &2010 - 2019 \\
Dong \cite{9378345} &BERT+LSTM &Stock data, Text &Stock Price &Tweet news, DJIA stocks &2019 - 2020 \\
Yang \cite{10.1145/3366423.3380128} &Transformer &Price, Text, Audio, Video &Volatility &S\&P 500 Earning Conference Calls dataset &2017 \\
Man  \cite{9442147} &BERT &Stock data, Text &Trading Strategy &CSI 300 index, Related news &2018 - 2019 \\
Hsu  \cite{hsu-etal-2021-semantics} &BERT,seq2seq &Text &Price Movement &Lap14,Rest14, Rest15, Rest16, SST-2, MR &2014 - 2016 \\
Yoo \cite{10.1145/3447548.3467297} &Transformer &Price &Price Movement &ACL181, KDD171, NDX100, CSI300, NI225, FTSE100 &2007 - 2019 \\
Ramos-Pérez \cite{ramos2021multi} &Multi‑Transformer &Price &Volatility &S\&P 500 &2016 - 2020 \\
Sonkiya \cite{sonkiya2021stock} &FinBERT, GAN &Stock data,Text &Stock Price &NYSE, NASDAQ, S\&P 500, NIffty 50, SSE, HKI, HKEX, News &2010 - 2020 \\
Li \cite{li_li_wang_jia_rui_2020} &BERT &Price, Text &Stock sentiment&Oriental Fortune online review, Monthly yield &2018 - 2020 \\
Li  \cite{li2017role} &Tensor Transformer &Price, Text &Stock movement &CSI 100 companies, news, social reviews &2010 - 2011 \\
Liu  \cite{liu-etal-2019-Transformer} &Transformer &Price, Text &Price Movement &S\&P 500 &2017 \\
Zhang \cite{ZHANG2022117239} &Transformer &Price, Text &Price Movement &Stocks, Twitter, CHRNN, Stocknet-dataset &2008 - 2019 \\
Colasanto  \cite{COLASANTO2022341} &AlBERTINO, Transformer &Price, Text &Stock Price &FinancialPhrasebank, EssilorLuxottica, Intesa SanPaolo, UnipolSai &2012 - 2022\\
Chen \cite{chen_chen_shen_shi_wang_zhang_2022} &Transformer &Price, Text &Stock movement &S\&P 500, Tweets, Stock consequential data &2014 - 2019 \\
Li \cite{li2022incorporating} &Transformer, LSTM &Price, Text &Stock movement &88 highest-ranked stocks & \\
Jiang \cite{jiang2017deep} &Deterministic Policy Gradient &Cryptocurrency price &Portfolio &Tradable cryptocurrency pair, cryptocurrencies &2014 - 2017 \\
Liang \cite{liang2018adversarial} &Adversarial PG &Price &Portfolio &China stock market and US stock market &-- \\
Li \cite{li2018learning} &A2C &Price &Policy &CSI 300 stocks &2005 - 2018 \\
Huang \cite{huang2018financial} &DQN &Tick-by-tick forex  &Policy &12 currency pairs from TrueFX.com &2012 - 2017 \\
Xiong  \cite{xiong2018practical} &DDPG &Price, Assets &Policy &30 DJIA stocks &2016 - 2018 \\
Yang \cite{yang2018investor} &IRL &Stock return rate &Policy, Stock Movement & SPX, IWD, IWC, SPY, DEM, VTI &2008 - 2015 \\
Chen \cite{chen2019application} &DQN &Price &Policy &S\&P 500 stocks &2000 - 2018 \\
Ponomarev \cite{ponomarev2019using} &A3C &Future Price &Policy &RTS Index futures (MOEX:RTSI) asks and bids &2015 - 2016 \\
Wang  \cite{wang2019alphastock} &Policy Gradient &Trading features &Portfolio &U.S. stock markers &1970 - 2016 \\
Bao  \cite{bao2019multi} &DDPG &Stock data &Policy &Simulation Environment &-- \\
Tsantekidis \cite{tsantekidis2020price} &PPO, DQN &forex price &Policy &Instrument combinations with currencies &2009 - 2018 \\
Liu  \cite{liu2020finrl} &DQN, PPO, TD3, DDPG, SAC, A2C &Balance Shares own price &Policy &NASDAQ-100, DJIA, S\&P 500, SSE 50, CSI 300, HSI, &- \\
Liu  \cite{liu2020adaptive} &iRDPG &Indicators &Policy &300 stocks on SSI and SZI &2016 - 2019 \\
Wang  \cite{wang2021deeptrader} &Policy Gradient &Indicators &Portfolio &DJIA 30, HSI 49, CSI100 &1971 - 2019 \\
Carta \cite{carta2021multi} &DQN &Price &Policy &S\&P 500 future market, J.P. Morgan, Microsoft &2012 - 2019 \\
Wang  \cite{wang2021commission} &REINFORCE, DDQN &Stock, States. &Portfolio &23 DJIA stocks, 23 SSE 50 Index stocks &2000 - 2018 \\
Briola  \cite{briola2021deep} &PPO, DQN &Volumns, Position &Policy &scale market activities description &2019 \\
Liu  \cite{liu2021finrl} &DQN, PPO, TD3, DDPG, SAC, A2C &States, Price, Sentiment &Policy &Dow-30, NASDAQ-100, S\&P-500, Cryptocurrencies, Currency, Futures &- \\
Li  \cite{li2021finrl} &Ensemble Strategy &States, Price, Sentiment & RL policy Acceleration. &NASDAQ stocks &2019 - 2021 \\
Liu \cite{liu2021finrlmeta} &DRL-based strategy &States, Price, Sentiment &RL environments &30 DJIA stocks, 10 market cap cryptocurrencies &- \\
Zhang \cite{zhang2018improving} &CMT &Stock data, Text &Stock Price &CSI 100, HKEX, Wind data and news, Guba posts &2015 \\
Duan \cite{duan2018learning} &Bi-LSTM &Stock Return &Stock Return &NYSE, Amex, NASDAQ news &2006 - 2015 \\
Xu \cite{xu2018stock} &Deep Generative &Text, Price &Stock Movement &Conglomerates sector stocks &2014 - 2016 \\
Sawhney \cite{sawhney2021quantitative} &RL &Tweets, News, Price &Stock Price &StockNet, China, Hong Kong &2014-2016 \\
Ang \cite{ang2022guided} &GNN &News, Price &Stock return &IN-NY, IN-NA, BE-NY, BE-NA &2015-2019 \\
Sawhney \cite{sawhney2021fast} &LSTM &Tweets, News, Price &Stock Price &StockNet, China, Hong Kong &2014-2016 \\
Zhou \cite{zhou2022stock} &GRU, CNN &Stock data &Stock Price &China &1991-2020 \\
Chandar \cite{chandar2022convolutional} &CNN &Stock data &Stock Trading &NASDAQ and NYSE &2009-2018 \\
Li \cite{li2022novel} &LSTM+GRU &Stock data + news &Stock Movement &S\&P 500 Index &2017-2018 \\
Wang \cite{WANG2022108285} &GCN &Stock data &Stock Trading &42 commonly used indices &2009-2021 \\
Duan \cite{duan2022factorvae} &VAE &Price &stock return &China A share &2017-2020 \\
Wei \cite{wei2019model} &Model-Based RL &LOB, Trade prints &Trading strategy &Hong Kong Stock Exchange. &2018 \\
Yu \cite{yu2019model} &Model-Based RL &Stock Price &Portfolio &U.S. equities &2005-2018 \\
Yin \cite{9533510} &GCN+GRU &Price, trade volume &Price, Price direction &DJIA,exchange-traded funds (ETFs) &2010-2020 \\
Zhao \cite{ZHAO2022117958} &BERT &stock price,text &Stock Movement &East Money &2020 \\
\end{longtable}
}
\end{center}
\section{Evaluation }\label{evaluation}

Evaluation measures play a crucial role in assessing the performance of stock market forecasting models. They are used to compare the forecasts made by different models to actual values. Commonly used evaluation metrics for classification models include accuracy-based metrics, while error-based metrics such as MAE and RMSE are commonly used for regression models. In this article, we classify the existing evaluation metrics into three categories: accuracy-based, error-based, and return-based. The accuracy-based and return-based metrics are considered to be better when their values are larger, while the error-based metrics are considered to be better when their values are smaller. Table \ref{tab3} provides a summary of the papers that use these three types of evaluation metrics.

\begin{table*}[!htp]\centering

\caption{The typical papers are in three evaluation methods, Accuracy-based, Error-based, and Return-based evaluation metrics.}\label{tab3}

\scriptsize
\begin{center}
\begin{tabular}{ p{2cm}|l|l|p{3cm}|p{1.2cm}|p{3cm} } 

\hline

Evaluation methods & Author & Year & Conference/Journal name & Methods & Evaluation methods \\

\hline

\multirow{6}{*}{Accuracy-based} &Nelson et al. \cite{nelson2017stock}&	2017&	IJCNN&	LSTM&	Accuracy, Precision, Recall, F1-score \\
&Zhao et al. \cite{zhao2017time}&	2017&	ICTAI&	LSTM&	Accuracy\\
&Hu et al. \cite{hu2018listening}&	2018&	WSDM&	GRU&	Accuracy \\
&Chen et al. \cite{chen2018incorporating}&	2018&	CIKM&	GCN&	Accuracy \\
&Deng et al. \cite{deng2019knowledge}&	2019&	World Wide Web Conference&	KDTCN&	Accuracy, F1-score\\
&Li et al. \cite{li2022incorporating}&	2022&	Complexity&	Transformer, LSTM&	Accuracy, MCC\\
\hline
\multirow{8}{*}{Error-based} & Rather, Agarwal and Sastry \cite{rather2015recurrent} & 2015 &	ESA&	RNN&	MSE, MAE\\ 
& Zhang et al. \cite{zhang2017stock}&	2017&	KDD&	RNN&	MSE\\ 
&Li et al. \cite{li2017role}&	2017&	MTA&	Tensor Transformer&	RMSE \\
& Li et al. \cite{li2018stock}&	2018&	PMLR&	LSTM&	MSE \\
&Feng et al. \cite{feng2019temporal}&	2019&	TOIS&	GCN&	MSE \\
&Rasheed et al. \cite{rasheed2020improving}&	2020&	IEEE&	CNN&	MAE, RMSE \\
&Eapen et al. \cite{eapen2019novel}&	2019&	IEEE&	CNN&	MSE \\
&Dong et al. \cite{9378345}&	2020&	IEEE&	BERT, LSTM&	RMSE \\
\hline
\multirow{8}{*}{Return-based} 
&Li Rao and Shi \cite{li2018learning}&	2018&	ISCID&	A2C&	SR\\
&Xiong et al. \cite{xiong2018practical}&	2018&	NeurIPS&	DDPG&	SR\\
&Feng et al. \cite{feng2019temporal}&	2019&	TOIS&	GCN &	MSE, IRR \\
&Sawhney et al. \cite{sawhney2020spatiotemporal} &	2020&	ICDM&	GCN&	SR \\
&Li et al. \cite{li2022graph}&	2022&	MTA&	GNN&	SR, IR, MD \\
&Zhou et al \cite{zhou2021trade}&	2021&	IJCNLP&	BERT&	Average Return \\
&Wang et al. \cite{wang2019alphastock}&	2019&	AAAI&	PG&	SR\\
&Wang et al. \cite{wang2021commission}&	2021&	AAAI&REINFORCE, DDQN &	AAR, SR\\

\hline

\end{tabular}

\end{center}
\end{table*}

\subsection{Accuracy-based Evaluation Metrics}
In this section, we will clarify a few key terms used throughout. The acronym TP stands for "True Positive," indicating a scenario where both the actual class and the model's prediction are positive. TN, or "True Negative," represents a situation where both the actual class and the model's prediction are negative. FP, or "False Positive," refers to when the model predicts a positive class, but the actual class is negative. Lastly, FN, or "False Negative," denotes when the actual class is positive, but the model's prediction is negative.
\\ 
\textbf{Accuracy.}
Accuracy, which evaluates the proportion of correctly classified predictions to the total number of predictions, is the most commonly used metric in classification tasks. It can be represented by the equation: $Accuracy = \frac{TP+TN}{TP+TN+FP+FN}$.

In a multi-classification confusion matrix, the correctly classified samples are represented on the diagonal line from top left to bottom right. This metric, known as Accuracy, evaluates the overall performance of a model in identifying samples. For instance, in the study conducted by \citet{hu2018listening}, Accuracy was used to measure the annualized rate of return, and in the research by \citet{chen2018incorporating}, it was employed to evaluate the outcomes of stock price prediction. However, it is important to note that Accuracy may not be an appropriate metric for datasets with imbalanced class distribution, and additional indicators should be employed to provide a more accurate representation of the model's performance. \\ 
\textbf{Precision, Recall, F-measure.}
The precision, recall and F1-score can be represented as equation $Precision = \frac{TP}{TP+FP}$, $Recall = \frac{TP}{TP+FN}$, and $F1 = \frac{2\times Precision\times Recall}{Precision+Recall} = \frac{2\times TP}{2\times TP+FP+FN}$.

In a multi-classification task, the performance for each class can be evaluated using precision, recall, and F1-score. Precision refers to the number of true positive cases that are predicted by the binary classifier. This metric reflects the reliability of the model in correctly identifying positive samples. Recall, on the other hand, measures the number of true positive cases in the test set that are predicted by the binary classifier. It indicates the model's ability to detect positive samples. The F1-score is a combination of both precision and recall, and it has been widely used in recent works such as \cite{deng2019knowledge, nelson2017stock, shen2012stock, 10.1145/3447548.3467297, COLASANTO2022341} to evaluate the performance of neural networks. The F1-score aims to strike a balance between precision and recall for a fair evaluation of the model. However, it should be noted that when the model's performance is evaluated using precision, recall, and F1-score, it may lead to an imbalance and ignore the true negative cases. Therefore, the Matthews Correlation Coefficient (MCC) can also be introduced as a possible evaluation strategy.

\noindent\textbf{Matthews Correlation Coefficient (MCC).}
The MCC (Matthews Correlation Coefficient) is a useful tool for evaluating the performance of a single-value classification model. It provides a summary of the information contained in a confusion matrix, which is a matrix used to represent the results of a classification model. The confusion matrix is typically represented in the following format:
$MCC = \frac{TP\times TN-FP\times FN}{\sqrt{(TP+FP)(TP+FN)(TN+FP)(TN+FN)}}.$
\\
The MCC is a measure of the correlation between predicted and actual samples, with a range of values from -1 to 1. A value of 1 indicates a perfect positive correlation, while a value of -1 indicates a perfect negative correlation, which occurs when the classifier misclassifies. The MCC is more robust to imbalanced categories compared to accuracy-based metrics. Several studies, such as \cite{li2022incorporating, xu2018stock, zhang2018improving, wang2021coupling}, have used MCC to evaluate their prediction results.

\subsection{Error-based Metrics}
Evaluating the performance of a prediction model can be done by comparing the predicted values to the actual values. One of the most widely used methods for this is measuring the error between the two. A lower error value indicates a better performance. In the context of stock market prediction, various error-based evaluation metrics are commonly utilized. These include the mean absolute error, mean square error, root-mean-square deviation, and mean absolute percentage error. In these metrics, a higher value indicates a better prediction.

\noindent\textbf{Mean Absolute Error (MAE).} The metric of MAE, which stands for Mean Absolute Error, calculates the average number of absolute differences between the predicted and actual values. Several studies, such as \citep{rather2015recurrent, rasheed2020improving}, have employed MAE as a means of evaluating the discrepancy between the actual and predicted values. The formula for MAE is as follows: $MAE = \dfrac{1}{n} {\sum{{i=1}^{n}} \lvert{y{i}-\hat{y_{i}}} \rvert}$.

\noindent\textbf{Mean Square Error (MSE).} The MSE, or Mean Squared Error, is a metric used to determine the average squared distance between the actual value and the predicted value. Unlike the MAE, or Mean Absolute Error, which calculates absolute errors, the MSE amplifies errors to their maximum by using the sum of squares of errors. Several studies, such as \citep{rather2015recurrent, zhang2017stock, li2017role, feng2019temporal, eapen2019novel}, have employed the MSE method to evaluate the discrepancy between actual and predicted values. The formula for MSE is represented as $MSE = \dfrac{1}{n} \sum{{i=1}^{n}} (y{i}-\hat{y_{i}})^{2}$

\noindent\textbf{Root-Mean-Square Deviation (RMSE).} The Root Mean Squared Error (RMSE) is a commonly used metric for evaluating the accuracy of predictions by measuring the square root of the second sample moment of the differences between predicted and actual values. It is similar to Mean Squared Error (MSE), with the only difference being the inclusion of the square root. It is therefore challenging to determine which evaluation metric is superior. In the works of \citep{li2017role, rasheed2020improving}, the authors employed RMSE to assess the discrepancy between predicted and actual values. The equation for RMSE is represented as: $RMSE = \sqrt{\frac{\sum_{i=1}^{n}(y_{i}-\hat{y_{i}})^{2}}{n}}$.

\noindent\textbf{Mean Absolute Percentage Error (MAPE).} The Mean Absolute Percentage Error (MAPE) is a metric used to evaluate the accuracy of forecasting models. It calculates the average of the absolute percentage difference between the predicted and actual values. This method has been widely used in the literature, as seen in studies such as \cite{zhou2022stock, ang2022guided}. The formula for MAPE can be represented as: $MAPE = \frac{100\%}{n}\sum_{i=1}^{n}\left | \frac{y_{i}-\hat{y_{i}}}{y_{i}} \right |$, where $n$ is the number of observations, $y_i$ is the actual value, and $\hat{y_i}$ is the predicted value.

\subsection{Return-based Evaluation Metrics}
Evaluating the accuracy of stock market predictions can be done effectively using return-based evaluation metrics. Two commonly used metrics in finance for assessing earnings are the return ratio and the Sharpe ratio. The higher the value of these metrics, the better the prediction.\\ 
\textbf{Investment Return Ratio (IRR).} IRR, or internal rate of return, is a metric used to measure the performance of an investment. It is calculated by determining the percentage difference between the value of an asset at the current time ($p_t$) and the value of the same asset at the previous time ($p_{t-1}$), divided by the previous value ($p_{t-1}$). Studies such as \citep{feng2019temporal, sawhney2021fast} have used IRR as an evaluation metric to assess the return ratio of investments. The equation for calculating IRR is $Returnratio = \frac{p_{t}-p_{t-1}} {p_{t-1}} \times 100\%$.\\ 
\textbf{Average Annual Return (AAR).} The Average Annual Return (AAR) is a metric that measures the historical average return of a mutual fund as a percentage. Unlike the Internal Rate of Return (IRR), AAR calculates returns on an annual basis. It is particularly useful for assessing the performance of investments over a prolonged period. In their study, Wang et al. \cite{wang2021clvsa} utilized a dataset spanning seven years from 2010 to 2017 and employed AAR as one of the evaluation methods to determine the average annual rate of return.\\ 
\textbf{Sharpe Ratio (SR).} The Sharpe Ratio (SR) takes into account both return and risk and calculates the average return per unit of volatility in relation to the risk-free rate \cite{Sharpe49}. This is represented by the equation: $SR = \frac{R_{t}-R_{f}}{\sigma}{\times 100\%}$, where $R_{t}$ represents the return, $R_{f}$ represents the risk-free rate, and $\sigma$ represents the standard deviation of the returns \cite{matsunaga2019exploring}. Several studies \citep{li2018learning, xiong2018practical, sawhney2020spatiotemporal, wang2021commission} have utilized the SR as an evaluation metric to assess the performance of return ratios.

\section{Future directions and open issues}\label{future}

The stock market prediction task has greatly contributed to the advancement of machine learning, particularly in the areas of natural language processing (NLP) and reinforcement learning (RL). However, there are still several potential research directions and open questions that need to be addressed in order to further improve and develop this field.\\
\textbf{Improving Generalization Ability For Stock Market Prediction.}
The ability of a machine learning model to accurately classify or predict unseen data is known as generalization. In the context of stock market prediction, a deep learning model must possess both high timelessness and strong generalization capabilities in order to be effective. However, some previous methods have struggled to generalize well to real-world trading scenarios or perform poorly on certain subsets of unseen data. Recent studies have suggested that incorporating self-supervised learning tasks into classification tasks can improve generalization, as demonstrated in works such as \cite{mohseni2020self, https://doi.org/10.48550/arxiv.2206.06606, hendrycks2019selfsupervised}. Further research in this area, both in terms of exploring existing methods and developing new ones, may be a promising direction for stock market prediction tasks. A self-supervised approach to enhancing generalization may be worth investigating in the future.\\

\noindent\textbf{Integrating Deep Learning Techniques with Online Learning Approaches}
Online learning is a training approach that utilizes the results of online training as feedback in order to optimize models. This method is particularly useful in mitigating the effects of volatility, uncertainty, and high noise factors in the stock market. Its application in stock market investment strategies is valuable as investors must constantly adjust their investment plans based on changes in stock prices. Online learning enables the simultaneous updating of the model and automatic control of the difference between predicted results and desired values. Other areas of application for online learning methods include dealing with abruptly changing time series. For example, Habibi \citep{prescott2007bayesian} proposed using a Bayesian setting for online change point detection, taking into account sudden variations in the Dow Jones Industrial Average's daily results. While this work contributes to detecting change points in time, it does not provide feedback for changing trading policies. We believe that integrating online learning and machine learning holds great potential for stock market prediction.\\

\noindent\textbf{Improving Evaluation and Datasets for Stock Market Prediction.}
Currently, many stock market prediction models only evaluate intermediate performance metrics such as stock movement prediction accuracy. However, it is unclear how well these models can support a practical trading system and there is a lack of uniform evaluation criteria for profitability. Each paper often uses different evaluation metrics on different datasets. Therefore, new stock market prediction models should be able to evaluate financial-relevant metrics, which can be grouped into three categories: profit criterion, including Annualized Rate of Return (ARR); risk criterion, including Maximum DrawDown (MDD) and Annualized Volatility (AVol); and risk-profit criterion, including Calmar Ratio (CR), Sortino ratio (SoR), and Annualized Sharpe Ratio (ASR). Furthermore, the stock market prediction task is currently fragmented with a lack of unified benchmark datasets and clear task descriptions, which greatly hinders the progress of this field.

\noindent\textbf{Improving Time Series Anomaly Detection for Stock Market Prediction.} It is a practical proposition to quickly and effectively identify out-of-performance stocks from among thousands of stocks across the market. Instability in financial markets poses significant risks to investors; examples of instability include market crashes due to systemic risk and abnormal stock price movements caused by artificially large publicity. The most common stock market prediction models failed to capture the best trading points without considering the existence of anomaly outliers. Time-serious anomaly detection will facilitate stock market prediction for capturing outliers in stock market trading prices, which can help investors adjust their strategies and reduce investment risk. In addition, the model can be used to model multiple financial time series datasets and capture anomalies in the companies of interest. To this end, a promising and essential anomaly detection would be to design a better mechanism based on time series anomaly detection tasks to capture the best trading points for prediction tasks when trading in the real world.

\noindent\textbf{Tasks All In One On Continual Learning For Stock Market Prediction.} Continual learning is a technique for training a model on multiple tasks consecutively while retaining information learned from previous tasks, even when the data from those tasks is no longer available. This enables neural networks to continuously accumulate knowledge and mitigate "catastrophic forgetting" in tasks such as stock prediction. However, current deep learning models for stock prediction are typically trained on static, homogeneously distributed data that cannot adapt or extend over time. To the best of our knowledge, there are no continual learning models specifically designed for stock market prediction, as the fluctuation of the stock market environment requires models to autonomously acquire new skills and adapt to new situations. Existing stock market prediction methods evaluate a single task on a single dataset, which can lead to overfitting of recent input data. Continual learning methods, such as those based on parameter isolation, can overcome this issue by freezing a portion of parameters after learning each task, allowing for more accurate and effective updating of the model for new tasks. 

\noindent\textbf{Leveraging Distributional RL For Stock Trading.}
Quantitative trading algorithms still struggle with balancing profit and risk due to the volatility and noise in the financial market \citep{an2022deep, sun2021Reinforcement}. One potential solution is the use of distributional RL, first proposed by Bellemare et al. in their paper C51 \cite{bellemare2017distributional}. Distributional RL goes beyond traditional values by utilizing a defined random variable, whose expectation represents the state-action value, to form the distributional Bellman equation \citep{bellemare2017distributional}. This equation is proven to be contracted under the measure of $p$-Wassertain distance. Many state-of-the-art Q-learning algorithms in RL are distributional, such as C51 \cite{bellemare2017distributional}, Quantile Regression DQN (QR-DQN) \cite{dabney2018distributional}, Implicit Quantile Network (IQN) \cite{dabney2018implicit}, and Fully Parameterized Quantile Function (FQF) \cite{yang2019fully}. Distributional RL can provide more information about the distribution of returns, which can help algorithms reduce risk or improve robustness. Previous research has demonstrated the effectiveness of Distributional RL in Atari games, where algorithms achieved scores higher than human players. However, there has been limited exploration of the application of Distributional RL in financial trading. Therefore, it is worth investigating the potential of distributional RL in this field.

\noindent\textbf{Treating Stock Trading As Partially Observable Markov Decision Process.}
RL algorithms have been widely used in financial trading, as discussed in section \label{RL}. These include model-free methods such as policy gradient, Q-learning, and hybrid methods. However, these methods assume a fully observed MDP which does not accurately reflect the open and ever-changing nature of the financial market. To address this issue, there are two potential solutions for future research.
The first solution is to collect all the transactions to make the dynamics fully observed, which may require significant storage and computational resources. For example, Briola et al. \cite{briola2021deep} have used transaction data on a small scale, which serves as a potential direction for future research. Another solution is to approximate the dynamics using a model-based RL approach. Researchers such as Wei et al. \cite{wei2019model}, Yu et al. \cite{yu2019model}, and Liu et al. \cite{liu2020adaptive} have demonstrated the effectiveness of transition dynamics models. Thus, the application of model-based methods in financial trading has a considerable potential and is worth exploring. By using a transition dynamics model, the policy could be capable of planning for a longer horizon.

\section{Conclusion}\label{conclusion}
In this paper, we present a comprehensive examination of the most prominent studies on utilizing deep learning for stock market prediction. To aid in understanding and organizing previous research in this area, we propose a classification system for categorizing and grouping similar works. Additionally, we provide an overview of current primary methods, evaluation metrics, and datasets used in stock market prediction. We also explore open questions and highlight promising future directions for machine learning research in stock market prediction. Through this survey, we aim to provide readers with a thorough understanding of the use of deep learning in stock market prediction.

\bibliographystyle{ACM-Reference-Format}
\bibliography{bibliography}
\end{document}